\documentclass[longauth]{aa}

\usepackage{graphicx}
\usepackage{hyperref}
\usepackage[varg]{txfonts}

\usepackage{caption}
\usepackage{subcaption} 
\usepackage{multicol}
\usepackage{multirow}
\usepackage{booktabs}

\usepackage{placeins}

\usepackage{mathtools} 
\DeclarePairedDelimiterXPP\BigOSI[2]%
  {\mathcal{O}}{(}{)}{}%
  {\SI{#1}{#2}}

\usepackage{float}
\usepackage{amsmath}
\usepackage{xcolor}

\usepackage{pifont} 

\usepackage{lipsum} 

\hypersetup{
  colorlinks=true,   
  urlcolor=blue,     
  linkcolor=blue,     
  citecolor=blue
}

\begin{document}

\title{First Statistical Study of Over 100 Magnified Stellar Events at Redshift $z \approx 0.725$ with JWST}

\author{J. M. Palencia    \thanks{palencia@ifca.es} \inst{1} \and
        Fengwu Sun \inst{2} \and
        J. M. Diego \inst{1} \and
        Yoshinobu Fudamoto \inst{3} \and
        Anton M. Koekemoer \inst{4} \and
        Christopher N. A. Willmer \inst{5} \and
        Eduardo Iani \inst{6} \and
        Xiaojing Lin \inst{7} \and
        Justin D. R. Pierel \inst{8} \and
        Alfred Amruth \inst{9,10} \and
        Tom Broadhurst \inst{11,12,13} \and
        W. Chen \inst{14} \and
        Liang Dai \inst{15} \and
        Daniel Espada \inst{16,17} \and
        Alexei V. Filippenko \inst{18} \and
        Seiji Fujimoto \inst{19,20} \and
        Patrick L. Kelly \inst{21} \and
        Mingyu Li \inst{7} \and
        Sung Kei Li \inst{9,10} \and
        Ashish Kumar Meena \inst{22} \and
        Jordi Miralda-Escudé \inst{23,24,25} \and
        P. Morilla \inst{11} \and
        Mitchell F. Struble \inst{26} \and
        Hayley Williams \inst{27} \and
        Rogier A. Windhorst \inst{27} \and
        E. Zackrisson \inst{28} \and
        Ruwen Zhou \inst{29} \and
        Adi Zitrin \inst{30}
        }

\institute{
              Instituto de Física de Cantabria (CSIC-UC), Avda. Los Castros s/n, 39005 Santander, Spain \and
              Center for Astrophysics | Harvard \& Smithsonian, 60 Garden St., Cambridge, MA 02138, USA \and
              Center for Frontier Science, Chiba University, 1-33 Yayoi-cho, Inage-ku, Chiba 263-8522, Japan \and
              Space Telescope Science Institute, 3700 San Martin Drive, Baltimore, MD 21218, USA \and
              Steward Observatory, University of Arizona, 933 N. Cherry Avenue, Tucson, AZ 85721, USA \and
              Institute of Science and Technology Austria (ISTA), Am Campus 1, 3400 Klosterneuburg, Austria \and
              Department of Astronomy, Tsinghua University, Beijing 100084, China \and
              Space Telescope Science Institute, 3700 San Martin Drive, Baltimore, MD 21218, USA \and
              Department of Physics, The University of Hong Kong, Pokfulam Road, Hong Kong \and
              The Hong Kong Institute for Astronomy and Astrophysics, The University of Hong Kong, Pokfulam Road, Hong Kong, P. R. China \and
              Department of Theoretical Physics, University of Basque Country UPV/EHU, Bilbao, Spain \and
              Ikerbasque, Basque Foundation for Science, Bilbao, Spain \and
              Donostia International Physics Center, Paseo Manuel de Lardizabal, 4, San Sebastián, 20018, Spain \and
              Department of Physics, Oklahoma State University, 145 Physical Sciences Building, Stillwater, OK 74078, USA \and
              Department of Physics, University of California, 366 Physics North MC 7300, Berkeley, CA 94720, USA \and
              Departamento de F\'{i}sica Te\'{o}rica y del Cosmos, Campus de Fuentenueva, Edificio Mecenas, Universidad de Granada, E-18071, Granada, Spain \and
              Instituto Carlos I de F\'{i}sica Te\'{o}rica y Computacional, Facultad de Ciencias, E-18071, Granada, Spain \and
              Department of Astronomy, University of California, Berkeley, CA 94720-3411, USA \and
              David A. Dunlap Department of Astronomy and Astrophysics, University of Toronto, 50 St. George Street, Toronto, Ontario, M5S 3H4, Canada \and
              Dunlap Institute for Astronomy and Astrophysics, 50 St. George Street, Toronto, Ontario, M5S 3H4, Canada \and
              Minnesota Institute for Astrophysics, University of Minnesota, 116 Church St. SE, Minneapolis, MN 55455, USA \and
              Department of Physics, Indian Institute of Science, Bengaluru 560012, India \and
              Institut de Ci\`encies del Cosmos (ICCUB), Universitat de Barcelona (UB), c. Martí i Franqu\`es, 1, 08028 Barcelona, Spain \and
              Departament de Física Qu\`antica i Astrofísica (FQA), Universitat de Barcelona (UB), 08028 Barcelona, Spain \and
              Institució Catalana de Recerca i Estudis Avançats (ICREA), 08010 Barcelona, Spain \and
              Department of Physics and Astronomy, University of Pennsylvania, 209 South 33rd Street, Philadelphia, PA 19104, USA \and
              School of Earth and Space Exploration, Arizona State University, Tempe, AZ 85287-6004, USA \and
              Observational Astrophysics, Department of Physics and Astronomy, Uppsala University, Box 516, SE-751 20 Uppsala, Sweden \and
              Tsung-Dao Lee Institute, Shanghai Jiao Tong University, 520 Shengrong Road, Shanghai 201210, People’s Republic of China \and
              Department of Physics, Ben-Gurion University of the Negev, P.O. Box 653, Beer-Sheva 8410501, Israel
              }

\abstract{Highly magnified 
stars at cosmological distances ($z \gtrsim 0.7$) become detectable thanks to microlensing by intracluster stars near the critical curves of galaxy clusters. Multi-epoch photometric campaigns targeting caustic crossing galaxies magnified by massive galaxy clusters enable the detection of these objects as transient events. Such stars provide unique opportunities to study stellar populations at early cosmic times, probe the nature of dark matter, reveal small-scale structure in the cluster, and improve lens models. To date, only a few dozen high-redshift stars have been reported, with a single lensed galaxy, the Dragon, holding the current record of 44 detections. These numbers, however, remain insufficient to exploit their full potential.
In this paper, owing to the inclusion of new observations, we report the identification of more than 100 magnified stellar events in the Dragon, behind the massive galaxy cluster Abell 370. The relatively low redshift of the Dragon ($z\approx0.725$) facilitates the detection of its most massive stars. Using imaging data from three different cycles (2022--2024) with the James Webb Space Telescope, we apply a time-domain technique to identify flux variations associated with caustic-crossing events. From the spatial distribution of stellar events we constrain the high-end slope of the stellar luminosity function, finding $\beta=2.18^{+0.20}_{-0.30}$. Alternatively, assuming a fixed slope, we constrain the microlens surface mass density. In addition, we examine the parity asymmetry of the detected caustic-crossing events, a proposed probe of wave dark matter, and find that it remains present. We also use the events to trace the regions of highest magnification, offering an alternative way to map the system critical curves.}


\maketitle
\section{Introduction} \label{sec:intro}
Small sources, such as individual stars within strongly lensed galaxies, can reach extremely large magnifications ($\mu\sim10^6$) when crossing the critical curves (CCs) \citep{Miralda-Escude1991}. When microlenses in the foreground lens are included, caustic crossing events (CCEs) become more frequent, typically producing lower but still substantial magnifications, often in the range of thousands \citep{Venumadhav2017,Kelly2018,Diego2018b,Oguri2018}. 
The combined action of the cluster- or galaxy-scale macrolens \citep{Li2025b} and the population of microlenses, typically stars but potentially also compact dark matter (DM) objects or density fluctuations associated with Fuzzy Dark Matter (FDM) \citep{Oguri2018,Diego2018b,Dai2018,Diego2024b,Palencia2024,Palencia2025,Broadhurst2025,Ji2025,Muller2025,Croon2025}, can produce total magnifications of several hundred to thousands \citep{Diego2018b,Diego2019,Palencia2024}. This enhancement can raise the observed flux of distant stars above the detection limit with a 1~hr observation using the James Webb Space Telescope (JWST; \citealt{Gardner2023}).

These highly magnified events are transient in nature. A highly magnified star spends most of its time in regions where the total magnification of multiple (unresolved) microimages is close to that predicted by the smooth macro model. However, as the star approaches a microcaustic, the magnification becomes dominated by the associated two brightest microimages and reaches a maximum when the microcaustic is crossed; after the crossing, these two brightest images disappear and the magnification is rapidly reduced back to its typical value (this sequence occurs in reverse in half the cases). The duration of these events depends on several factors, including the detection threshold of the telescope, the relative source motion, the intrinsic brightness and size of the star, and the geometry of the lens-source system. Observed events typically exhibit characteristic timescales ranging from days to weeks, but the biggest stars can be observed for months. In addition, luminous stars that are close to the cluster smooth critical curve may remain detectable at their typical magnification in medium exposures with the Hubble Space Telescope (HST) and JWST. 

When multi-epoch observations of the same lensed galaxy are available, a fraction of stars is expected, on average, to be undergoing a CCE at any given epoch. Some of these events can be detected in sufficiently deep  observations as transients. Between two epochs, stars that are highly magnified at the first epoch will generally return to their macromodel magnification at the second epoch, and vice versa. Comparing fluxes between epochs therefore reveals large-amplitude variability associated with these transient magnification peaks \citep{Rodney2018,Chen2019,Kaurov2019,Diego2022,Kelly2022,Welch2022,Diego2023,Meena2023a,Meena2023b,Yan2023,Fudamoto2025}. Given the large number of stars within lensed arcs, many stars are expected to experience CCEs, although only the most luminous ones remain detectable at current observational depths (magnitude $m_{\rm AB} \approx 28$--30). Caustic-crossing galaxies at lower redshift benefit from a smaller luminosity distance and are therefore easier to detect.

Typical multi-epoch observations of lensed galaxies yield only a handful of CCEs per gravitational arc. However, the increased photometric depth of JWST observations has notably increased these numbers. The current record holder is a strongly lensed galaxy at redshift $z \approx 0.725$, commonly referred to as ``the Dragon,'' with 44 events detected in the difference images from two epochs \citep{Fudamoto2025}. The Dragon arc is among the lowest redshift and highest surface brightness caustic-crossing galaxies  known~\citep{Soucail1988,Smail1991,Kelly2022}, and is intersected by powerful cluster caustics multiple times, thus making it the ideal target to detect microlensing events.

The Dragon arc was the first giant arc ever identified \citep{Soucail1987,Paczynski1987}, and it is lensed by the massive galaxy cluster Abell~370, one of the six clusters targeted by the Hubble Frontier Fields (HFF) programme \citep{Lotz2017}. Deep HFF imaging, together with the Beyond Ultra-deep Frontier Fields and Legacy Observations (BUFFALO) survey \citep{Steinhardt2020} and Multi Unit Spectroscopic Explorer (MUSE) spectroscopy \citep{Bacon2010}, has enabled the identification of $\sim 40$ spectroscopically confirmed multiply lensed systems, which are used to construct accurate lens models \citep{Lagattuta2017,Lagattuta2019,Lagattuta2022,Diego2018a,Diego2025,Eid2025,Niemiec2023}.

Single-event light curves can provide valuable information about the microlenses responsible for these events. The duration of an event depends, among other factors, on the mass of the microlens, while multiple caustic crossings may reveal information about the internal structure of the lens, the relative tangential velocity between the lensed galaxy and cluster caustic (complementary to the radial velocity from spectroscopic measurements), or constrain certain DM models. Compact DM candidates, such as primordial black holes, behave as point-like lenses and in most configurations (minima and maxima points) produce a single microcaustic when combined with the macro potential. In contrast, extended dark objects, such as ultra-compact axion minihalos, can generate additional sets of microcaustics \citep{Croon2025}. If sufficiently long light curves are available, they can also constrain the overall population of microlenses, potentially allowing different DM models to be distinguished. For instance, the standard Peccei–Quinn, post-inflation axion that solves the strong CP problem predicts dark matter minihalos that can perturb the microlensing light curves at CCEs \citep{Dai2020}. In the case of ultra-light axions, motivated by string theory, wave effects become significant on astrophysical scales, leading to coherent density fluctuations that can enhance or reduce the underlying macromagnification. In such cases, a modulation in the light curves may appear, with characteristic spatial separations related to the de Broglie wavelength, which depends inversely on the axion mass~\citep{Broadhurst2025}. This behaviour extends to other bosonic candidates, as particles with masses below $\sim1\,\mathrm{eV}$ can exhibit wave-like dark matter phenomenology, producing matter over- and underdensities on scales that increase with decreasing mass.

While event light curves are highly informative, their construction requires high-cadence follow-up observations, making them observationally demanding. An alternative approach relies on multi-epoch analyses that identify the positions of different events and  measure changes in magnification between epochs. The spatial distribution of events alone can already provide important information about the microlens population, the location and structure of the system CCs, and the presence of nearby structures in the form of millilenses \citep{Palencia2025,Diego2022,Diego2024b}.

Beyond studies of the microlensing population, the identification of individual stars, particularly when spectra are available, enables novel constraints on  stellar populations at early cosmic times. In particular, since only the most luminous stars can be detected through this technique, one can study the high end of the stellar luminosity function, or the initial mass function (IMF). This is not accessible with classic photometric or spectroscopic studies of distant galaxies since here the observed flux is often dominated by the less luminous (or less massive) stars.  
Owing to their immense distances, these stars are observed as they were when the Universe was only a fraction of its current age, providing a unique opportunity to test models of stellar and galaxy evolution across cosmic time. To date, a candidate for the most distant star is Earendel \citep{Welch2022}, at $z=6.2$, corresponding to a cosmic age of $\sim 900$~Myr, although some recent work suggests this might actually be a stellar cluster~\citep{Pascale2025}. At similar or higher redshifts, metal-free stars may also become detectable through microlensing \citep{Windhorst2018,Zackrisson2024}.

When enough data are available for a single arc, stellar-population analyses that focus on the brightest stars can be performed. JWST observations are more sensitive to red supergiant stars (RSGs) due to their longer wavelength coverage, while HST, operating at shorter wavelengths, preferentially detects blue supergiants (BSGs) \citep{li2025c}. Nevertheless, JWST remains sensitive to a broad range of stellar types, including many BSGs, except for the hottest and bluest sources. Other candidates, such as yellow supergiants, including Cepheids, can potentially be detected in deep JWST observations of low-redshift caustic-crossing galaxies \citep{Diego2024a}. When a sufficiently large number of individual stars is detected within the same arc, statistical analyses of the stellar population become possible --- for example, the slope of the luminosity function, or alternatively the IMF, can be constrained \citep{Li2025a,Williamsh2025,Meena2025}. A shallower IMF would produce a larger number of massive stars, leading to a higher number of detectable events and allowing events to occur farther from the CC, where the magnification is lower. Other population parameters, such as metallicity, dust attenuation, or star-formation history, can also be better constrained when transient events are included. For instance, higher metallicities produce cooler, redder stars, favouring detections in longer-wavelength filters, while older populations with little recent star formation would alter the ratio of BSGs to RSGs owing to their different lifetimes \citep{li2025c}. These events therefore act as an additional constraint that can help break degeneracies commonly encountered in standard photometric or spectroscopic analyses of galaxies.

In this paper, we analyse the Dragon arc ($z=0.725$) \citep{Soucail1987}, lensed by the galaxy cluster Abell~370 ($z=0.375$). We perform a multi-epoch analysis using four JWST/NIRCam filters spanning three years, searching for transient events. We identify a total of 104 distinct events associated with the same galaxy, or $\sim 2.5$ times the value reported in earlier studies, owing to the inclusion of additional data. This significantly expanded sample opens the door to further analyses of the stellar population within the Dragon arc, with implications for both the nature of DM and the properties of the IMF at intermediate redshifts. We also study the spatial distribution of events to probe the small-scale structure of the CCs around the arc and to test the slope of the luminosity function. Finally, we provide a catalogue containing the position of each star detected in the arc\footnote{The catalogue is available at \href{https://doi.org/10.5281/zenodo.19439926}{Zenodo}.}.

This paper is organised as follows. Section~\ref{sec:data} describes the adopted JWST/NIRCam data. Section~\ref{sec:methodology} presents the methodology employed to detect transient events, including point-spread-function (PSF) reconstruction, residual image processing, and detection strategies. In Section~\ref{sec:results}, we present the catalogue of microlensed stars and the events detected outside the arc. We discuss the results and their implications in Section~\ref{sec:discussion}, and summarise our conclusions in Section~\ref{sec:conclusion}. Throughout this paper, we assume a standard flat cosmological model with $\Omega_m=0.3$, $\Omega_\Lambda=0.7$, and $h=0.7$. With this cosmology, at the redshift of the lens ($z_l=0.375$), one arcsecond corresponds to 5.16~kpc, and to 6.44~kpc at the redshift of the Dragon ($z_s=0.725$). We use the prefixes macro-, milli-, and micro- to refer to different lensing mass scales. The first corresponds to the cluster-scale lens, with typical masses of $M\sim\mathcal{O}(10^{15},\mathrm{M}\odot)$. The second refers to subhalo-like structures, with masses of $M\sim\mathcal{O}(10^{6}$–$10^{9}\,\mathrm{M}\odot)$, while the latter denotes the small perturbers responsible for the CCEs, such as stars, stellar remnants, or compact DM structures, with masses of $M\sim\mathcal{O}(10^{-1}$–$10^{2}\,\mathrm{M}_\odot)$.

\begin{figure*} 
  \includegraphics[width=\linewidth]{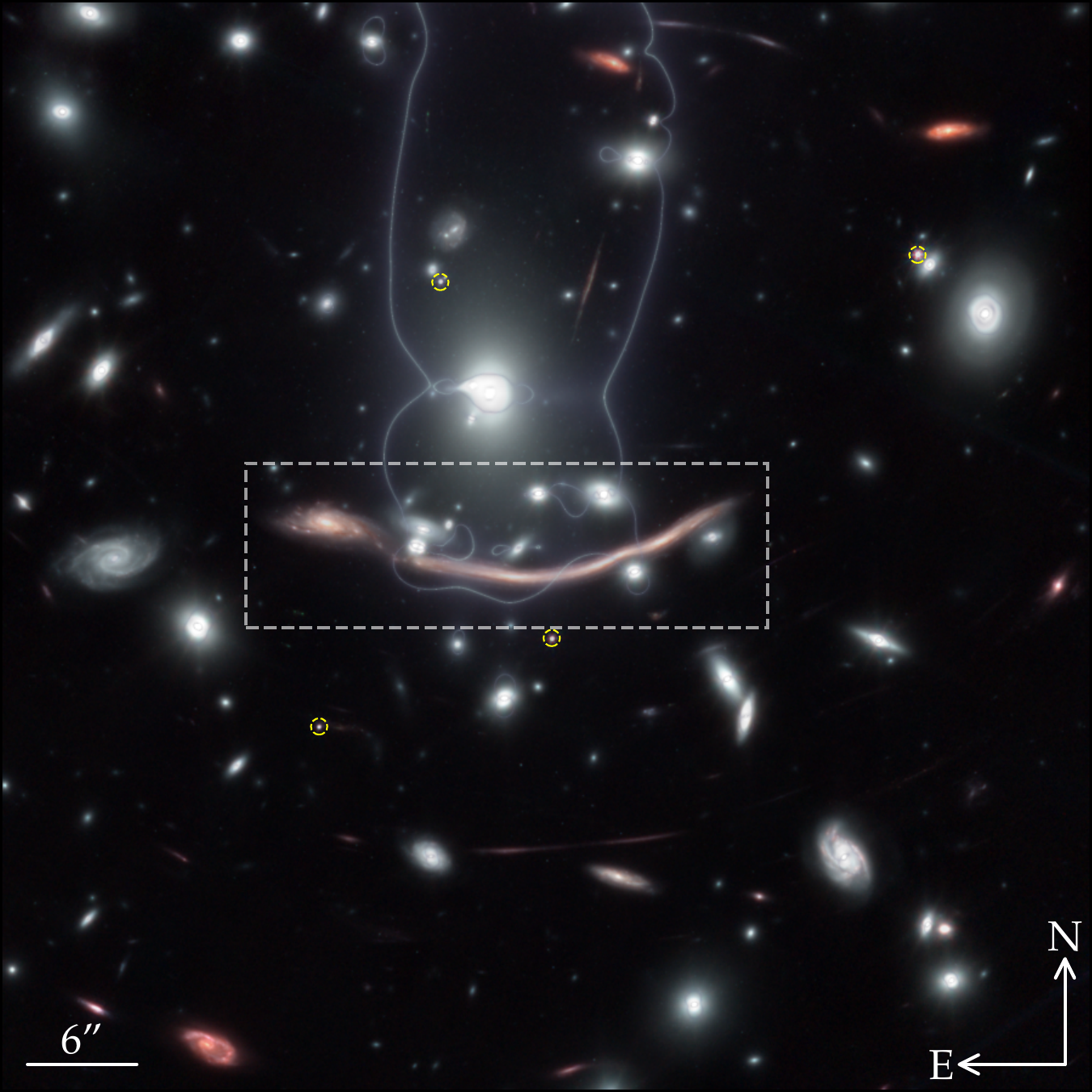} 
      \caption{Cutout region of Abell~370 ($60'' \times 60''$) centred on the Dragon arc. Composite image built from the NIRCam filters F410M (red), F200W (green), and F150W (blue), covering a wavelength range from 1.3 to 4.3~$\mu$m. The white dashed rectangle marks the region of interest where we search for transient events. Overlaid are the critical curves from the lens model for Abell~370 from \citep{Diego2025}, where the white contours indicate the absolute magnification. Yellow dashed circles show the positions of the subset of stars within this field of view used for the PSF characterisation.}
         \label{fig:abel370}
\end{figure*}
\section{Data} \label{sec:data}
We make use of deep JWST images obtained with NIRCam, combining all publicly available observations acquired before January 15, 2025, in the F150W, F182M, F200W, and F210M filter bands. We do not use images obtained at longer wavelengths because their coarser angular resolution hinders our ability to detect CCEs reliably, therefore increasing the likelihood of spurious detections. 

All JWST/NIRCam \citep{RiekeM2023} imaging data products used in this work have been drizzled onto a common World Coordinate System (WCS) frame, with a scale of $0.03''$~pixel$^{-1}$. 
We focus on a small region of $29'' \times 9''$ around the Dragon arc; for reference, we show a composite colour image of $60'' \times 60''$ including this area, with the CC at $z_s=0.725$ from the lens model of \citep{Diego2025}, in Fig.~\ref{fig:abel370}. All subsequent analyses are performed in this common astrometric frame to ensure consistency across epochs. 

The JWST mosaics were produced 
using a modified version of the JWST pipeline (v1.11.2), similar to that used by \citet{Fudamoto2025} for the same target. Customised processing steps include, but are not limited to, corrections for 1/$f$ noise, wisp removal, and modelled background subtraction. The dataset combines observations from multiple programmes, including Cycle~1 GTO-1208 CANUCS (PI: Willott; \citealt{Doyon2023,Willott2022}, Cycle~2 GO-3538 (PI: Iani; \citealt{Iani2023}), GO-2883 MAGNIF (PI: Sun; \citealt{FuS2025}), and Cycle~3 GO-5324 (PI: Pierel; \citealt{Pierel2024}). A summary of the observations used can be found in Table~\ref{tab:observations_summary}.
This uniform treatment enables a direct comparison between JWST observations and multi-epoch analyses of transient events.

We adopt the recent lens model of \citet{Diego2025} as our reference model. It is built with the free-form code WSLAP+ \citep{Diego2005,Diego2007}, a hybrid method that allows for the combination of weak and strong lensing constraints. For this particular model, only strong lensing constraints were used, particularly the image positions of 44 multiply lensed families. Tens of individual constraints are also taken for the Dragon arc itself from multiply lensed clumps in the arc. For further details on the lens model we refer the reader to \citet{Diego2025}.

\section{Methodology} \label{sec:methodology}
The goal of this paper is to extract a reliable catalogue of transient events within the Dragon arc produced by stellar microlensing. The detectability of such events relies on stars whose differential magnification between different epochs exceeds a threshold above the noise level and whose intrinsic brightness, when combined with the lensing magnification, is above the observational depth in at least one epoch. Under these conditions, a proper differential analysis reveals flux variations large enough to be considered as a robust detection.

For this analysis, an accurate characterisation of instrumental effects at each epoch is essential, including the PSFs which vary in orientation according to the corresponding position angles (PAs). Equally important is a robust characterisation of the baseline level of flux fluctuations along the arc, which allows us to quantify the significance of each event and to avoid misclassifying intrinsic flux variations of the source galaxy as microlensing events.

In this section, we first describe the construction of the PSFs required for the subtraction analysis. We then present the preparation of the residual images, the characterisation of the baseline fluctuation levels, the different event-detection algorithms employed, and finally the classification of the detected events.

\subsection{Empirical PSFs and data preparation}
A differential analysis relies on a pixel-by-pixel subtraction of fluxes from epoch $i$ to epoch $j$. With this subtraction, we can then proceed to search for contiguous pixels with the expected spatial shape given by the PSF. In an ideal scenario the subtraction of images taken in two epochs would involve obtaining the two images with the same exact PA. However, in practice, observations are often made with different PAs, and difference images show residuals around bright compact sources, even if the flux of the source has not changed between observations. To account for this effect at a pixel level, we convolved the $i$-th epoch with the PSF of the $j$-th epoch and vice versa. This results in the PSF acting twice over each epoch, but the effective PSF is now identical in both epochs. In addition, all candidate events are visually inspected to remove spurious detections, such as pixel-level artefacts (e.g. hot pixels) that may appear point-like after convolution. 

To obtain the empirical PSF (ePSF) models of each epoch and band, we select isolated, unsaturated stars within the field of view. 
The number of satisfactory stars selected per filter and epoch varies from 3 to 14, typically ranging from 19 to 23 AB mag~\citep{Oke1983}. 
Once a set of stars is located, we build an ePSF model through sigma-clipped stacking using the {\tt Photutils}~\citep{Bradley2024} \texttt{EPSFBuilder} pipeline.
All ePSF models are oversampled by a factor of 2, and we have visually inspected the selected stars and obtained models for quality assurance.


\begin{table}
\caption{Summary of JWST/NIRCam observations used here.}
\label{tab:observations_summary}
\centering
\begin{tabular}{l c c c}
\hline\hline
Programme & Date & Filter(s) & Exposure (h) \\
\hline

GTO-1208 
& 2022-12-29 
& F150W, F200W 
& 1.78 \\

\hline

\multirow{2}{*}{GO-5324} 
& 2024-09-09 
& F150W, F200W 
& 0.21, 0.26 \\
& 2024-12-17 
& F150W, F200W 
& 0.41  \\

\hline

\multirow{3}{*}{GO-3538} 
& 2023-12-19 
& F182M, F210M 
& 2.70  \\
& 2023-12-20 
& F182M 
& 1.29 \\
&2023-12-21
& F210M 
& 1.29 \\
& 2023-12-22 
& F182M, F210M 
& 1.41, 1.32 \\

\hline

\multirow{3}{*}{GO-2883} 
& 2024-07-25 
& F210M 
& 1.29 \\
& 2024-07-31 
& F182M 
& 2.58 \\
& 2024-08-02 
& F210M 
& 1.29 \\
\hline

\end{tabular}
\end{table}

\subsection{Residual analysis}
Once the ePSFs have been characterised for each filter and epoch, we proceed with the residual analysis. A straightforward approach consists of subtracting two arbitrary epochs, each image being convolved with the ePSF of the other epoch. This procedure has been successfully used in previous work \citep{Fudamoto2025}. However, given the larger number of epochs per filter included in this study, we adopt a slightly different strategy.

For a given epoch of interest (epoch~$i$), we construct a reference image as the median of the remaining epochs,
\begin{equation}\label{eq:median_intensity}
\bar{I}_i(x) = \mathrm{median}\left(\{ I_j(x) \}_{j=0,\,j\ne i}^{N}\right),
\end{equation}
where $I_j(x)$ is the intensity at pixel $x$ for observing epoch~$j$, and $\bar{I}(x)$ represents the median intensity of all epochs except epoch~$i$ at the same pixel. This reference image provides a cleaner representation of the arc where transient events are averaged out. At pixels where no transient occurs, the median intensity remains stable up to noise fluctuations, while at positions where a transient is present in any of the averaged images, its contribution is strongly suppressed, as it is much less likely to appear in more than one epoch. As a result, such fluctuations are driven toward the underlying baseline level.

This averaging process, however, slightly increases the effective noise level of the reference image compared to the deepest images, since shallower observations contribute to the overall noise. Consequently, some very faint events that could in principle be identified in a direct epoch-by-epoch subtraction may fall below the detection threshold in this combined analysis.

Once the reference image is constructed, we apply the same procedure to the ePSFs. 
This definition of the reference image, which we take to represent the background state of the arc in the absence of microlensing transients (or excitations if we follow the quantum analogy), offers several advantages. First, averaging reduces the impact of noise, thereby lowering the expected number of false positives. Second, the significance of detected events becomes more robust, unless the event is present in all epochs. Finally, for a large number of epochs within the same filter, this approach substantially reduces the complexity of the analysis.



A drawback of this method is a slight reduction in depth, such that some faint events detectable with the standard approach may be missed. However, most false positives also occur at these low flux levels. Given the increasing availability of multi-epoch observations, we therefore consider that the advantages of this methodology significantly outweigh its limitations, and we adopt it for this work and for future analyses.

After this procedure is completed, we convolve epoch~$i$ with the resulting ePSF for the stacked image (built as shown in Eq.~\ref{eq:median_intensity}), and vice versa with the stacked image with the $i$-th ePSF. Then, we simply subtract one image from the other to obtain the differential image for epoch~$i$. Once the differential image is produced, we use {\tt SExtractor}~\citep{Bertin1996} to estimate and remove the background level associated with any large-scale background variation. For this task we adopt the same parameters as \citet{Fudamoto2025}. The remainder is the residual image we will use for the search of transients.

Next, we define a set of masks required for the analysis. These include both selection and exclusion regions: the Dragon arc itself, used to restrict the search for events, nearby bright foreground objects that must be excluded (particularly those overlapping with the arc), and the Dragon's ``head,'' a near complete image of the galaxy mapped into the Dragon arc, located close to the arc that exhibits a similar surface brightness. This extra image is used as an estimator of the intrinsic flux fluctuations of the source galaxy. The Dragon arc mask is therefore not used to remove the arc, but to define the region where events are searched for, as our selection criteria are calibrated using the source galaxy itself, through the Dragon's head. Transient events outside this region can, and have (see, for instance, event $\alpha$ in \citep{Fudamoto2025}), be detected, but their significance would not be meaningful.

We also identify three images of the galactic bulge within the arc. Two of these correspond to the central region of the galaxy, 
where the photon noise 
is large and significantly hinders the detection of microlensing events. Moreover, even after the convolution step, residuals remain at the location of bright point sources such as the nucleus, owing to imperfect PSF modelling and photon noise. We therefore exclude these regions, as they would otherwise give rise to large intrinsic fluctuations that could be misidentified as CCEs. The third image corresponds to the surrounding regions of the bulge. In this case, some events can still be detected; however, their significance is evaluated not with respect to the global galaxy fluctuations (excluding the bulge), but relative to a local region surrounding the bulge in the Dragon's head. 

For the construction of the masks we make use of {\tt SExtractor}~\citep{Bertin1996}. We select the CANUCS 1.78~hr exposure in the F200W band for this task, as it corresponds to the deepest epoch and provides the highest signal-to-noise ratio (S/N). For the background estimation, we adopt a box size of $70 \times 70$ pixels, together with a background filter with a width and height of 3 pixels. Once the background is estimated, it is subtracted from the image, and objects are detected using a S/N threshold of 3. This threshold is taken as the standard deviation of the residual image after masking the Dragon arc, its head, and the bright foreground sources. For both the object extraction and the background estimation, all remaining parameters are kept at their default values in {\tt SExtractor}. 

We then use the resulting segmentation map and retain the regions corresponding to cluster members. For the Dragon's head, we construct an elliptical mask with the best-fitting parameters returned by {\tt SExtractor}. Using this same elliptical shape, we scale the radii down by a factor of $\sim 3$ to define the near-bulge region mask, and by an additional factor of 3 to define the bulge mask that we exclude from the analysis (see Fig.~\ref{fig:masks}) .

Within the Dragon's head, we identify a small number of bright features, likely globular clusters or compact star-forming regions within the galaxy, whose residuals are accentuated in the residual maps owing to imperfect PSF modelling and photon noise. These regions are masked using circular apertures with radii of 3 pixels. In this way, the Dragon's head's mask or the near-bulge mask, after excluding both the bulge and these substructures, can be used to calibrate the significance of flux fluctuations in the Dragon arc, given their similar surface brightness.

Finally, the bulge images along the Dragon arc are also masked using elliptical apertures that follow the local orientation of the arc at the position of each image. Their spatial extent is chosen such that the bright fluctuations observed in the residual maps are fully covered. An image of the Dragon arc illustrating all the masks used in this analysis is shown in Fig.~\ref{fig:masks}.

\subsection{Identification of microlensing events}
Transients appear as localised flux variations along the arc. In the microlensing regime, the multiple micro-images produced during a CCE are not spatially resolved. Therefore, the characteristic extent of the flux variation lies well below the pixel scale. As a consequence, these events are imaged with a morphology that closely follows the PSF and can, to first order, be approximated as Gaussian. Since these events are typically only slightly above the detection limit, the contribution from the PSF wings (e.g., diffraction spikes) is expected to be negligible, remaining below the noise level. Given the fact that their spatial signature is known, the problem reduces to identifying clusters of contiguous pixels consistent with this profile and exceeding a given amplitude threshold. The threshold is chosen such that the detected candidates can be confidently associated with genuine CCEs (or other compact transients), rather than to noise fluctuations or intrinsic surface-brightness variations of the source.

\begin{figure*} 
\includegraphics[width=\linewidth]{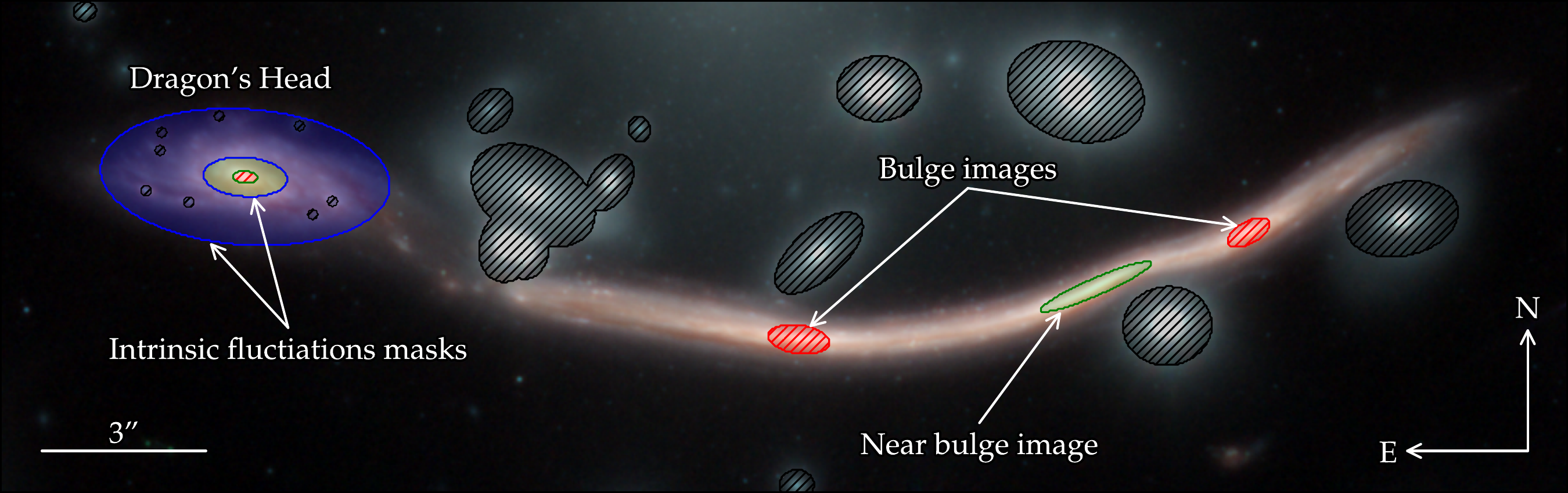} 
\caption{Same as Fig.~\ref{fig:abel370}, but for the $29'' \times 9''$ region around the Dragon arc. Shaded areas indicate the different masks used in our analysis. Hatched regions are masked out: bulge images (red), cluster members and substructures (black). Non-hatched areas define the regions of interest: green regions mark the near-bulge area, while the blue region traces the rest of the Dragon's head; the fluctuation distribution in the Dragon's head is used to characterise events in the tail.}
         \label{fig:masks}
\end{figure*}

The threshold selection is guided by an empirical estimate of the intrinsic fluctuation level. We exploit the availability of the Dragon's head, which acts as an extra image with a similar surface brightness, but without microlensing events (because it is far away from the CC), to measure the distribution of flux variations in the residual maps in the absence of microlensing, and adopt it as a baseline for the Dragon arc. 
Therefore, the observed residual fluctuations provide a proxy for intrinsic variability within the source galaxy. Some very compact and bright regions, such as globular clusters and the bulge, exhibit larger fluctuations in the difference images, potentially exceeding those expected from CCEs. We thus restrict the analysis to the masks defined in the previous subsection (see Fig.~\ref{fig:masks}, blue and green shaded regions on the northeast side), excluding these non representative areas. We adopt two masks in order to better sample the increased fluctuation level near the bulge, and to use this information to characterise events detected in the central part of the arc (rightmost green shaded region in Fig.~\ref{fig:masks}).

\begin{figure} 
  \includegraphics[width=\linewidth]{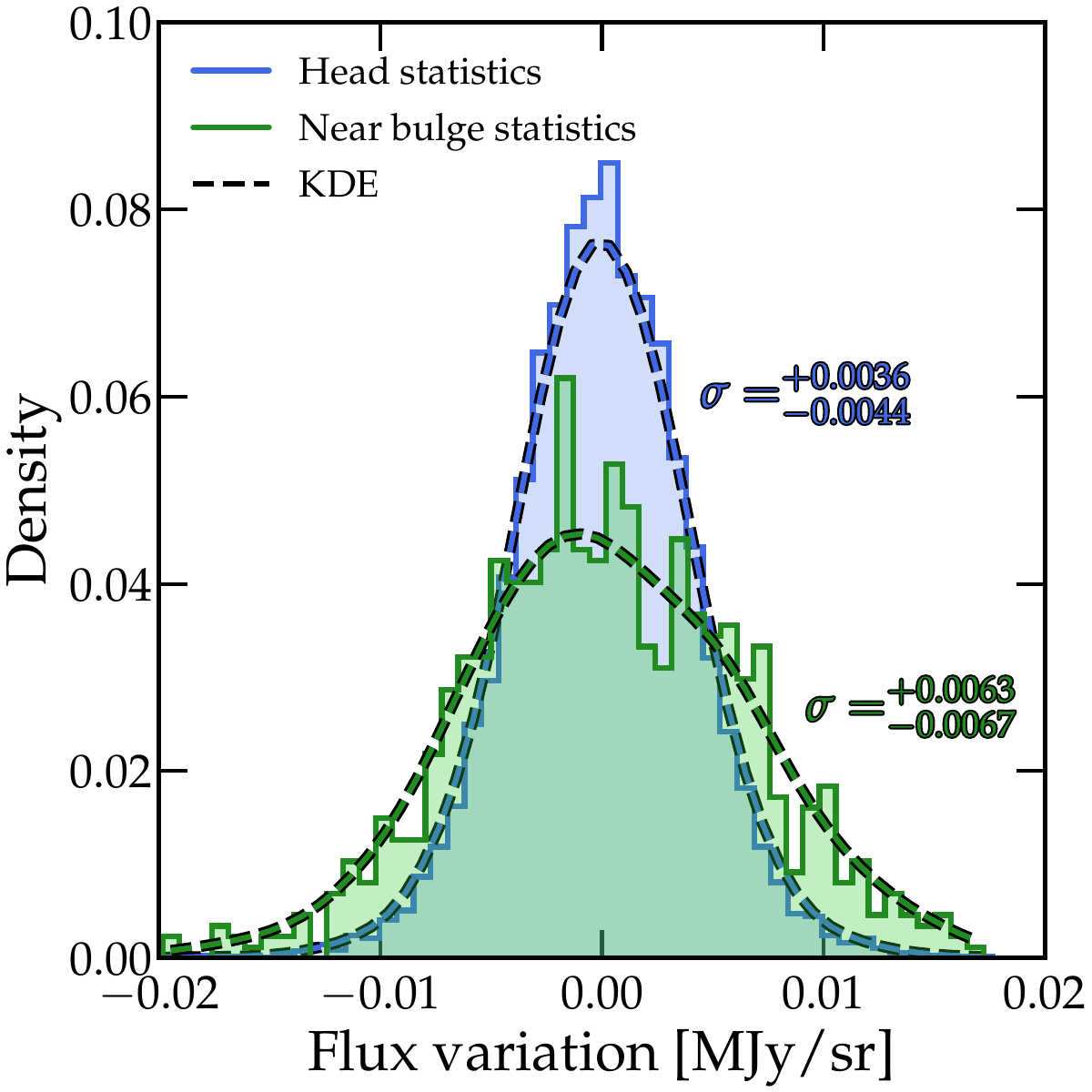} 
\caption{Pixel-wise flux fluctuation statistics within the Dragon's head, measured inside the predefined masks, for the F200W filter in the CANUCS~\citep{Willott2022} epoch. The blue and green distributions correspond to the Dragon's head and near-bulge region, respectively, with dashed curves showing their KDEs. The fluctuation amplitudes at which the integrated probability reaches the $1\sigma$ level are indicated.}
         \label{fig:head_stats}
\end{figure}

To characterise the intrinsic fluctuations, we first build a histogram of pixel values within the selected masks. We do not do this directly on the raw residual map, but instead on a version of the residuals convolved with a Gaussian kernel having a full width at half-maximum intensity (FWHM) of 2.5 pixels. This step is intended to amplify the information contained in contiguous pixels with the expected Gaussian morphology. We specifically follow the Gaussian kernel definition adopted by {\tt DAOFIND} \citep{Stetson1987}. The resulting histograms are close to Gaussian, centred at $\sim 0$, but with some kurtosis. To estimate the probability density function (PDF) of fluctuations robustly, we apply a Gaussian kernel density estimation (KDE)~\citep{Scott1992} and determine the pixel values corresponding to different $\sigma$ levels. This provides the basis for assigning a significance to CCE candidates and other transients. An example of this procedure for the F200W band and the CANUCS epoch \citep{Willott2022} is shown in Fig.~\ref{fig:head_stats}.

We now describe the two search algorithms employed in this work, namely {\tt DAOFIND}~\citep{Stetson1987}, particularly its implementation within {\tt photutils}~\citep{Bradley2024}, and {\tt SExtractor}~\citep{Bertin1996}. The former is our primary detection engine, as it is particularly efficient at identifying circular Gaussian sources, which is the expected morphology of CCEs in the residual images. The latter complements this approach by detecting sources with more complex morphologies, such as blended events or events embedded in regions of enhanced noise, whose spatial signatures deviate from a simple circular Gaussian profile but still exceed our detection threshold.

\subsubsection{DAOFIND}

{\tt DAOFIND} identifies local maxima above a user-defined threshold in an image convolved with a Gaussian kernel. We adopt the same kernel used to characterise the flux fluctuation statistics, namely a circular Gaussian with FWHM = 2.5 pixels. In the convolved image, the algorithm searches for pixels corresponding to local maxima with amplitudes exceeding the $5\sigma$ threshold derived from the flux variation PDFs measured in the Dragon's head.

This $5\sigma$ threshold 
is applied separately to the general arc and to the near-bulge region. As a result, fluctuations of similar amplitude in these two regions can be assigned different confidence levels, properly accounting for the local background properties. For each detected source, {\tt DAOFIND} returns the pixel coordinates, which we convert into RA and Dec using the common WCS. We then measure the peak residual value at the position of each event and estimate its significance using the noise PDFs derived from the Dragon's head.

For events where the flux variation appears as a local minimum, we repeat the analysis using the residual images with inverted sign.

\subsubsection{SExtractor}

{\tt SExtractor} detects objects in astronomical images by estimating the background and identifying pixels that satisfy a given threshold criterion. Detected objects must consist of a minimum number of contiguous pixels, specified by the user. The algorithm also performs efficient deblending of overlapping sources and provides photometric and geometric measurements for each detection. In addition, it supports PSF estimation and image filtering.

In our application, we require a minimum of four connected pixels with values exceeding $2.5\sigma$, where $\sigma$ is derived from the flux variation PDFs measured in the Dragon's head. To deblend crowded sources, we use a contrast parameter of 0.0005. Under the assumption that individual pixel values are statistically independent, the significance of an object can be written as
\begin{equation}\label{eq:significance_eq}
S_{\mathrm{obj}} = \sqrt{\sum_{i=1}^{n_{\mathrm{pix}}} S_i^2},
\end{equation}
where $S_i$ is the significance of the $i$-th pixel. For a Gaussian distribution, $S_i$ is defined as the pixel value $A_i$ divided by the standard deviation $\sigma$. For non-Gaussian distributions such as those considered here, we define an equivalent Gaussian significance as the $\sigma$ value of a Gaussian distribution that yields the same integrated probability above $A_i$ as our empirical distribution.

Applying Eq.~\ref{eq:significance_eq} to our detection criterion of four pixels greater than or equal to $2.5\sigma$ results in a nominal object significance of $5\sigma$ or larger. We note, however, that this assumption overestimates the true significance, as pixel values are not statistically independent. The signal in different pixels originates from the same event, in addition to the PSF effects, and the noise is also correlated due to PSF effects and the convolution applied during filtering. A full rigorous treatment would require computing the covariance matrix for each detected object, which is impractical given that the positions and relative amplitudes of the background sources are not known a priori. Although the noise covariance could in principle be estimated through random realisations of object-sized masks placed in source-free regions, we adopt a simpler empirical approach.

We compare the significances obtained from {\tt SExtractor} with those derived from {\tt DAOFIND} for sources detected by both methods. By construction, many of the objects identified by {\tt DAOFIND} also satisfy the {\tt SExtractor} detection criteria of four connected pixels above $2.5\sigma$. This allows us to directly assess the bias in the {\tt SExtractor}-based significance estimates on a source-by-source basis. We find that the bias is approximately constant across sources, with {\tt SExtractor} significances being overestimated by a factor of $\sim 2$. We therefore correct the significances of objects detected only by {\tt SExtractor} by dividing their nominal significance by this factor.

Finally, we repeat the entire procedure on the residual images with inverted sign in order to identify CCEs that appear as negative flux fluctuations.

\begin{figure*} 
  \includegraphics[width=\linewidth]{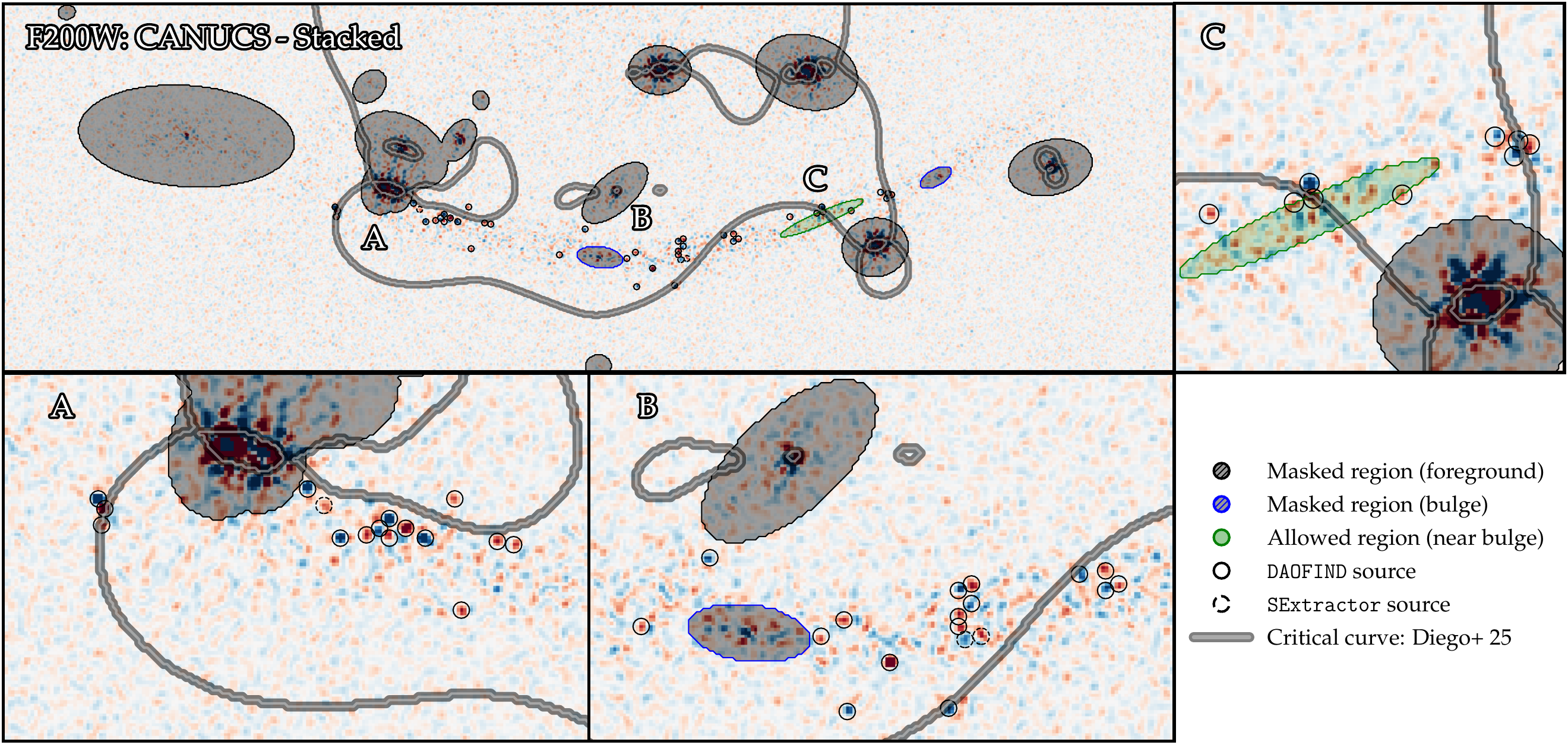} 
\caption{Transient events ($>5\sigma$ relative to fluctuations in the Dragon's head) detected in the Dragon arc from the CANUCS F200W residual image. We identify 43 events, comparable to the 44 reported by \citet{Fudamoto2025}. 
Black solid and dashed circles show detections from {\tt DAOFIND} and {\tt SExtractor}, respectively. The grey line traces the CC from the lens model of \citet{Diego2025}.}
         \label{fig:events}
\end{figure*}

\section{Results} \label{sec:results}
We first report the results for the CANUCS epoch in the F200W band. This epoch was previously analysed by \citet{Fudamoto2025}, who presented the largest collection of CCEs detected in a single galaxy to date. In their work, a pseudo-F200W image was constructed by combining the F182M and F210M images from the GO-3538 programme. In addition, they adopted an epoch-minus-epoch approach, whereas in this work we subtract a stacked image built from the median flux (Eq.\ref{eq:median_intensity}) of the remaining epochs. Furthermore, we employ a detection threshold calibrated directly from the same image, based on the flux variations measured in the Dragon's head, while \citet{Fudamoto2025} assumed a fixed threshold of $5\sigma = 0.038$~MJy~sr$^{-1}$. In our analysis, we also distinguish the significance of detections in the near-bulge region, where intrinsic flux fluctuations are larger, and apply a more stringent threshold in this area, which can naturally reduce the number of detected events close to the bulge. Finally, their {\tt SExtractor} detections did not include a correction for the assumption of independent pixels, although each detection was subsequently inspected and some unreliable sources were discarded.

By comparing our results with theirs, we are able to validate both the methodological changes introduced by our analysis and the robustness of their pseudo-filter constructed from the combination of the two medium-band filters. In our analysis, as shown in Fig.~\ref{fig:events}, we identify 43 events, compared to the 44 events reported by \citet{Fudamoto2025}. This difference of a single event, corresponding to $\sim2\%$, is in good agreement and supports the consistency of the two approaches. Note that we compare only the total number of events. Since the second epoch used in \citet{Fudamoto2025} is not the same as our median-stacked image, some events are expected to differ, and the counts may vary slightly. {\tt DAOFIND} events are shown as solid circles in Fig.~\ref{fig:events}, while {\tt SExtractor}-only events, a minority, are shown as dashed circles.

Once the methodology has been validated against previous results, we repeat the full analysis for the remaining F200W epochs and extend it to all available epochs in the other bands (F150W, F182M, and F210M). The final catalogue, including the reported events, their positions on the sky, and their associated significance, is presented in Table~\ref{tab:catalogue}. The catalogue is also provided in an accompanying {\tt astropy} table and a machine-readable ASCII table.

When applying the analysis to all observational epochs, we obtain a total of 224 detections. This number, however, includes multiple detections of the same physical event, either observed in different bands at the same epoch or persisting long enough to be detected across multiple epochs. A final version of the catalogue therefore requires filtering to account for these repetitions.

We first identify detections that coincide spatially on the sky, imposing a minimum separation of 2.5 pixels ($\sim0.075''$). When this criterion is met, we examine their observing epoch and detection band. If multiple detections occur at the same epoch, they may correspond either to a single star detected in multiple filters or to distinct stars undergoing CCEs at the same pixel position. As these two scenarios cannot be reliably distinguished without a dedicated spectral energy distribution (SED) analysis, and ideally spectra (which is beyond the scope of this work, and we do not have the required data), such detections are grouped and recorded as a single event.

Alternatively, detections may occur at the same position but at different epochs. If an observing epoch with no detection lies between two such detections, they are interpreted as independent events and recorded separately. These sources are of particular interest since they could correspond to intrinsically periodic sources, for instance Cepheid variable stars, which recent work predicts  should be detectable in the Dragon arc with JWST \citep{Diego2024a}. From the table we identify several candidates that appear in different epochs separated by several months, consistent with repeating long-period Cepheids (affected by time dilation), which are the ones expected to be detected in the Dragon arc. However, the number of epochs and cadence are not ideal to reliably identify Cepheid candidates, a situation that would be resolved with new data.  If, instead, the detections occur in consecutive epochs, it is not possible to determine whether they correspond to two distinct events or to the same microlensing event whose flux changes but remains above the detection threshold in both epochs. In this case, we conservatively record them as a single event.

\begin{figure*} 
  \includegraphics[width=\linewidth]{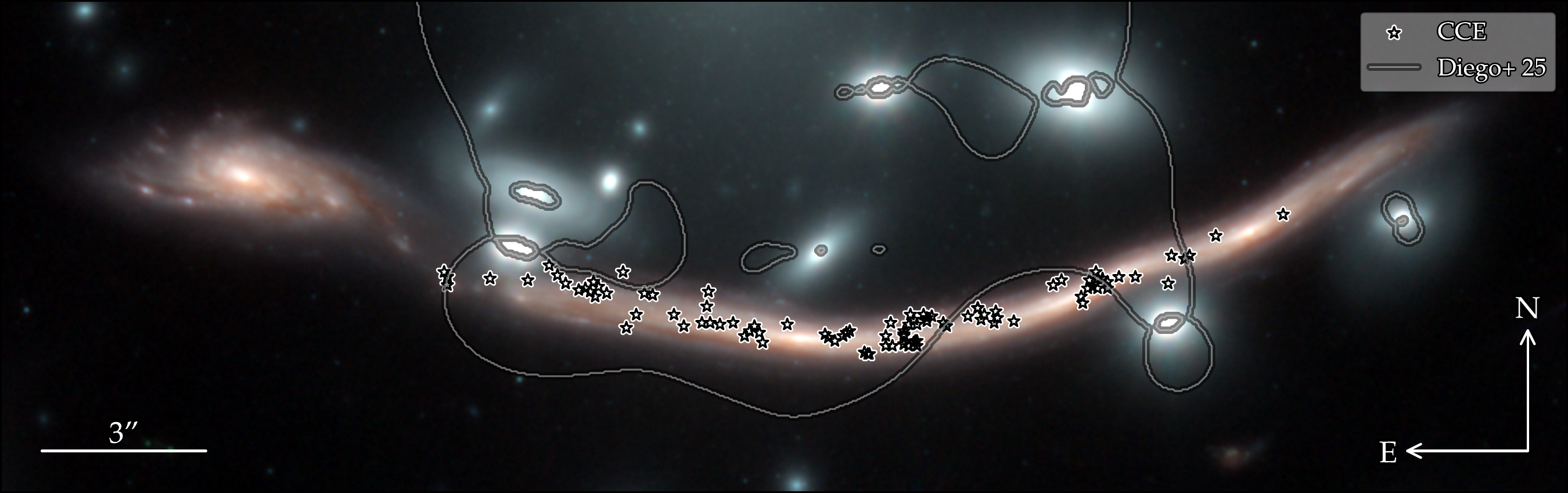} 
\caption{Locations of CCEs across the Dragon arc. The 104 unique events listed in Table~\ref{tab:catalogue} are indicated by star symbols. The best-fitting lens model CC from \citet{Diego2025} is overlaid as a grey line. The colour-composite image was constructed as described in Fig.~\ref{fig:abel370}.}
         \label{fig:events_over_dragon}
\end{figure*}

After applying the selection and filtering techniques described above, we identify 104 unique candidates. Some of these candidates are detected in consecutive epochs, resulting in up to 140 recorded detections including repeated events. All these events, together with basic information on the detection method, sky position, detection significances in the observed filters, and dates of observation, are listed in Table~\ref{tab:catalogue} in Appendix~\ref{app:catalogue}, and their spatial distribution over the Dragon arc is shown in Fig.~\ref{fig:events_over_dragon}

\begin{figure*} 
  \includegraphics[width=\linewidth]{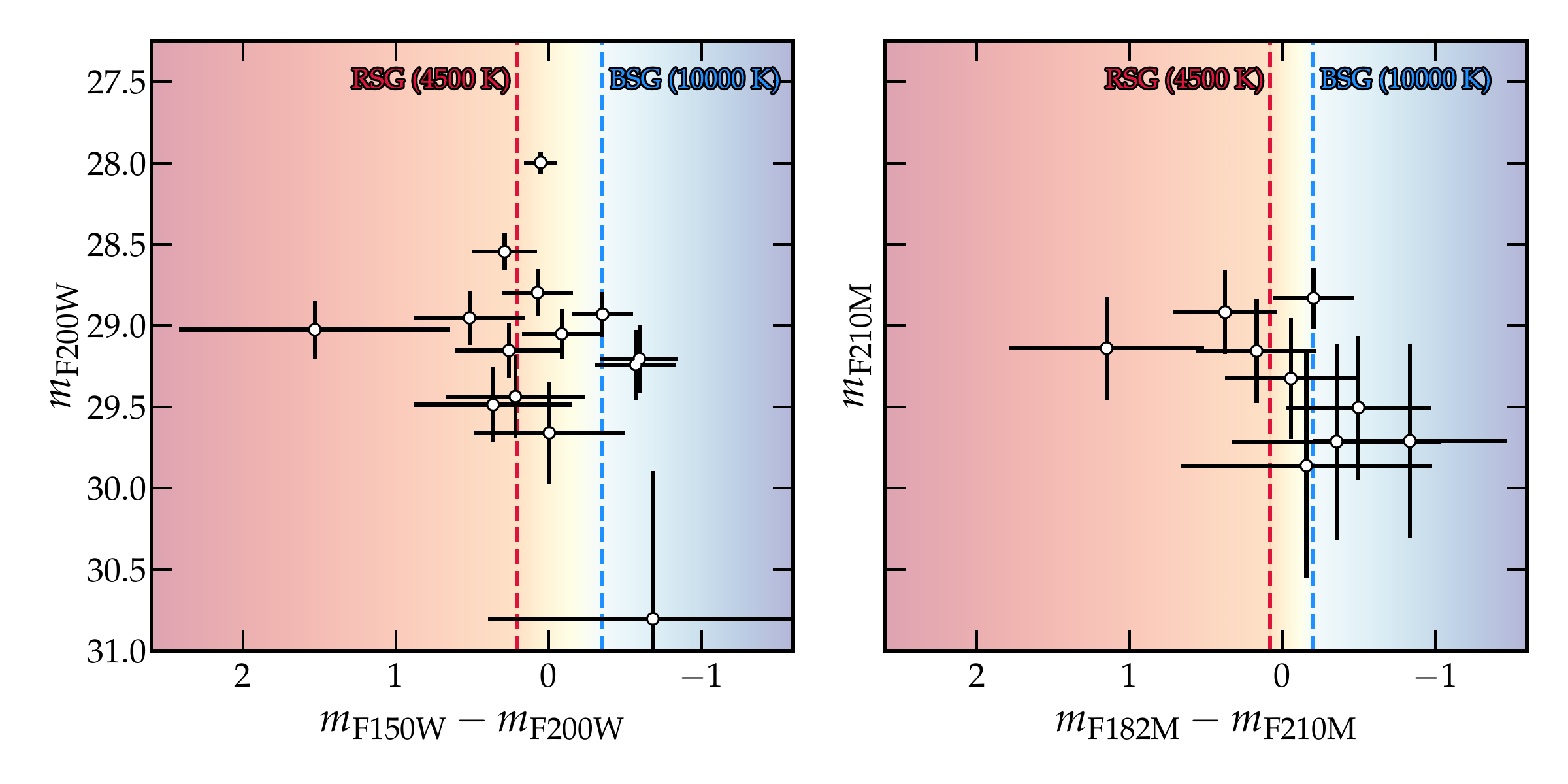} 
\caption{Colour–magnitude diagram for events detected in two filter bands within the same epoch, requiring $>5\sigma$ significance in both. 
Plotted magnitudes are values above the residual level after lensing magnification. Red and blue lines show the expected colours for a 4500~K RSG and a 10,000~K BSG, respectively, at the source redshift, using stellar spectra from \citet{Coelho2014}.}
         \label{fig:colour_magnitude}
\end{figure*}

Several consecutive detections, such as those associated with ID~44, are separated by only a few days and  therefore likely produced by the same star. Other cases, such as ID~32, appear first in 2022 and again one year later. In these cases, where only two filters are available for each detection epoch (F150W and F200W in 2022, and F182M and F210M in 2023), it is hard to distinguish between a single star remaining above the detection threshold for an extended period and two distinct stars located at the same pixel position undergoing separate CCEs at different times. Colour information (Fig. \ref{fig:colour_magnitude}) can, in principle, be used to break this degeneracy if the observed differences cannot be explained by the SED of an expected stellar source, such as a young, bright star (e.g. RSGs, BSGs, OB-type stars, or objects with similar absolute magnitudes). Given the uncertainty in the photometric measurements, the fact that additional effects such as dust reddening have not been taken into account, and the inability to measure magnification variations between epochs under the assumption that the same star is being observed, we have not used the colour information to clarify whether these events originate from the same source.

\section{Discussion} \label{sec:discussion}
In this section we review the properties of the events found in the Dragon arc, as well as the potential science that can be done with a large population of microlensed stars. We briefly describe the insights from stellar populations at high redshifts, the potential constraints on different DM models, and improvements to the lens model across the arc.

\subsection{Insights from stellar populations at $z\approx0.725$}
The background stellar population determines, to a great extent, the number of expected detectable CCEs. The lens model, both macro and micro, plays a key role in both the number of events and their spatial distribution, but for a fixed model, the number of events, their colour, spatial distribution, and even their duration depend on the background stellar population. The detection of individual stars through extragalactic microlensing opens the door to tight constraints on ancient massive stellar populations, which are hardly accessible through standard photometric analyses that are susceptible to degeneracies in the stellar population modelling. Ultimately, a combined fit, mixing information from microlensed events, spectroscopic and photometric analyses, offers the best approach toward a bias-free characterisation of the background stellar population~\citep{li2025c}. So far, this problem has been addressed only in the inverse way, where spectrophotometric analyses have been produced to obtain a guess of the stellar population, and then, through lensing, one can obtain a prediction of the expected events over the arc~\citep{Palencia2025,Li2025a,Williamsh2025}. But now, with the large number of events in the Dragon arc, we are in a position where we can start exploiting this optimal method, where we can compare the predicted stellar events and their types with different stellar population model predictions.


We have performed forced aperture photometry on the event locations in the residual maps at their epoch observations, for those events that have at least a $5\sigma$ detection in two filters. The aperture radius was 2 pixels ($0.62''$), and photometric errors were obtained by taking the standard deviation of 30 random apertures located on the northwest portion of the arc that shows no events and similar surface brightness. The resulting colour-magnitude diagram for the flux variations is shown in Fig.~\ref{fig:colour_magnitude}. These events are compatible with a blue-to-red ratio of 1, with a few of them showing a yellow colour ($\sim 6000$~K), as expected from Cepheids or yellow supergiants. Although we do appreciate a tendency for red stars, a blue-to-red ratio points in the direction of a relatively young stellar population or a recent star-formation episode~\citep{Garcia2010}. This ratio can also be linked to a shallower IMF, in which massive (intrinsically bluer) stars are more abundant~\citep{li2025c}.

The blue-to-red ratio may, of course, also be affected by interstellar dust reddening within the Dragon arc, and this is not something we can correct for on a star-by-star basis unless each event has multi-band photometry, as shown by \citet{Williams2025}. However, \citet{Patricio19} constrained the reddening within the Dragon arc to be $E(B-V)\lesssim 0.4$, based on MUSE data and under the assumption of a \citet{Calzetti00} dust attenuation law. Dust reddening at this level would shift the $m_\mathrm{F150W}-m_\mathrm{F200W}$ colour of stars by $\lesssim 0.3$ mag in Figure 6, which is smaller than the colour difference ($\approx 0.5$ mag) between $T_\mathrm{eff}=4500$ K and 10000 K stars. Hence, while dust attenuation may bias the observed blue-to-red ratio slightly in favour of red stars compared to the intrinsic distribution, it seems insufficient to move, for instance, all $T_\mathrm{eff}<4500$ K candidates to colours above the $T_\mathrm{eff}>10000$ K line. 

\subsection{Constraints on DM}
In addition to constraints on the stellar population, the number of events can be used to constrain the lens model, specifically the microlens models and possible millilenses, both linked to DM. As was shown by \citet{Palencia2025} for arcs with crowded foreground microlensing populations such as the Dragon, adding microlenses in the form of compact DM, such as primordial black holes, would not change the total integrated number of events. However, if the increase in microlenses as compact DM is large, the spatial distribution will change, predicting more events farther from the CC and reducing the events in the regions close to the CC, owing to the ``more is less'' effect~\citep{Diego2018b,Palencia2025}.

Regarding millilenses, these will locally boost the magnification around them (about 1 pixel), resulting in a clustering of events originating from the same pixel \citep{Diego2024b,Palencia2025}. The abundance of repeated events makes this possibility tantalising, but more data (and in particular more colour information) is needed. 
An alternative approach to reveal millilenses would come from light curves, events showing long-lasting magnification, especially in regions of low expected macromagnification, such as Mothra~\citep{Diego2023}, or events at positions where the macromagnification would otherwise be too low to expect CCEs, such as Godzilla~\citep{Diego2022}.

Finally, an interesting opportunity regarding constraints on DM is the parity asymmetry in the number of events. \citet{Broadhurst2025} showed that, under a FDM scenario, the surface-density fluctuations favour a larger number of CCEs on the negative-parity side of the CC with respect to the positive parity. A mild asymmetry is, however, also expected in the standard scenario, where the negative parity is intrinsically associated with a higher probability of extreme magnifications~\citep{Palencia2024}. In this context, FDM predicts a significantly enhanced asymmetry; therefore, the detection of a pronounced imbalance between parities could provide evidence in favour of a wave-like nature of DM. To test whether this scenario holds when adding the new events, we have created a density map by smoothing a map of zeroes, filled with ones at the pixel positions where events were found. To smooth this map, we have used a kernel $\mathcal{D}(r)$ to convolve the aforementioned map. The kernel can be written as
\begin{equation}
    \mathcal{D}(r) = \frac{1}{(r+r_{\mathrm{min}})^\alpha},
\end{equation}
where $r$ is the distance in pixels, and $r_{\mathrm{min}}$ and $\alpha$ are parameters that control the width and decay of the kernel, respectively. We found that values of $r_{\mathrm{min}}=5$ pixels and $\alpha=1$ represent well the spatial distribution while maintaining the local information of the events found. This is similar to Shepard's algorithm~\citep{Shepard1968} where usually $\alpha$ takes the value 2. Now, assuming the lens model by \citet{Diego2025}, we integrate the event density value over all the pixels whose macro-magnifications, according to our lens model, belong to a set of macromagnification bins, for both the positive and negative parities. Each macromagnification bin is assigned a normalised event density, $\mathcal{N}(\tilde{\mu})$, defined as
\begin{equation}
\mathcal{N}(\tilde{\mu}) =
\frac{1}{N}
\sum_{i=1}^{n_{\mathrm{pix}}}
\Sigma_{\mathrm{ev}}(i)\,
\Theta\!\left(\mu(i) \in \tilde{\mu} \pm \frac{\Delta\mu}{2}\right),
\end{equation}
where $N$ is the number of pixels whose magnification $\mu(i)$ lies within the bin interval $\tilde{\mu} \pm \Delta\mu/2$, with $\Delta\mu$ the bin width and $\tilde{\mu}$ the central value of the bin, chosen in logarithmic space. The quantity $\Sigma_{\mathrm{ev}}(i)$ denotes the event surface density at the $i$-th pixel. The function $\Theta$ equals unity if the condition inside the parentheses is satisfied and zero otherwise.

\begin{figure} 
  \includegraphics[width=\linewidth]{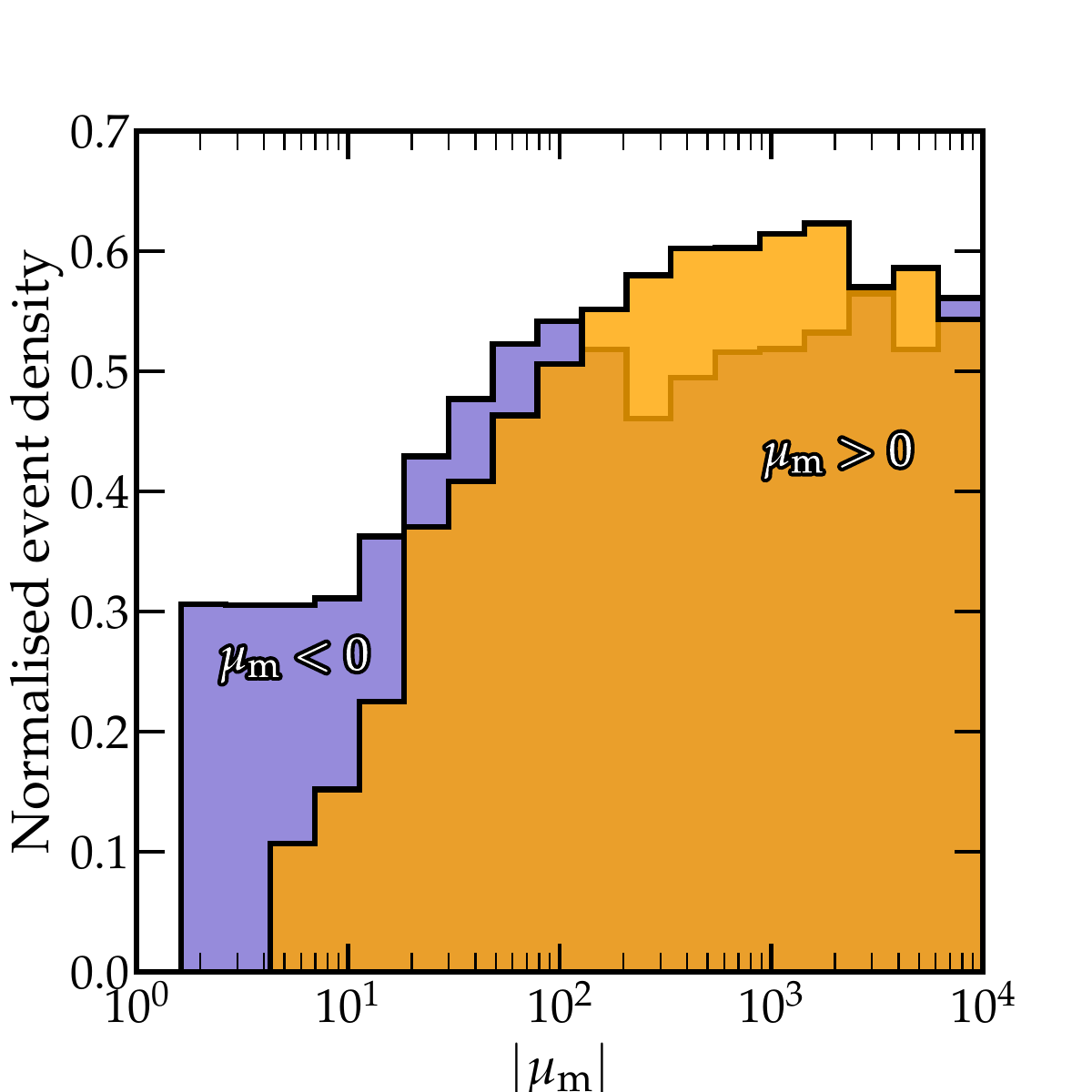} 
\caption{Normalised distribution of event density per macromagnification values. Orange bins show the positive macro-parity scenario, while the purple ones show the negative macro-parity counterpart. The macromagnification adopted here corresponds to the fiducial model of \citet{Diego2025}.}
         \label{fig:density_per_macromu}
\end{figure}

Such results are shown in Fig.~\ref{fig:density_per_macromu}. We observe that the events decay at similar rates for both positive and negative parities for low to intermediate macromagnification values, while the number of events drops for the positive parity near the CC and remains approximately constant at the CC for the negative parity. The negative parity also shows a bump at intermediate macromagnification values. This could provide a hint of a fuzzy nature of DM, although it may also arise from an incorrect location of the CC in the lens model. This test is inherently model dependent and therefore performs best in systems where the uncertainty in the CC position is well constrained, such as Warhol~\citep{Palencia2025,Williamsh2025}, for which most models converge on a consistent CC location. In such cases, more robust constraints on the nature of DM can be derived. Empirical CCs derived from symmetry arguments can also be used as model-independent constraints, such as those presented in \citet{Broadhurst2025}. In the latter case, the asymmetry is still present with the addition of new data.

In contrast, for the Dragon arc, although the number of events is sufficiently large for the statistics to be meaningful, current lens models exhibit significant dispersion in the predicted CC crossing of the arc. More stringent constraints on the CC position are therefore required before the observed parity asymmetry can be reliably used to test the nature of DM.

Ignoring the very high magnification regions (which are model dependent), and focussing on the lower magnification regions, the most interesting result from this figure is the clear increase in the number of events with magnification. If the luminosity function of the most massive stars, $dn/dL \propto L^{-\beta}$, has slope $\beta=2$, \citet{Diego2024b} show how the distribution of events should be uniform along the arc (near and far from the CC). Hence, the larger areas with smaller magnification should contain more events. 
We note that this result is closely related to the well-known behaviour in magnification space, where $\beta=2$ yields a flat distribution of images above a fixed flux threshold per unit $\log \mu$ in the high-magnification limit \citep[e.g.][]{Blandford1992}. However, finite source size effects introduce additional complexity at the highest magnifications, as the maximum amplification depends on the stellar radius and the local microcaustic structure. As a result, deviations from this simple behaviour are expected in practice, particularly near the critical curves.
The location of events shown in Fig.~\ref{fig:events_over_dragon} clusters near the CCs, suggesting that the system does not lie in the $\beta=2$ regime, and is instead more consistent with $\beta>2$. In this case we expect more events near the areas of greater magnification, in agreement with the results in Fig.~\ref{fig:events_over_dragon}. Assuming the \citet{Diego2025} lens model, we found from the number of events in the region near the CC ($\mu >100$) and in the region far from it ($\mu <100$) that the slope of the LF is $\beta=2.18^{+0.20}_{-0.30}$, similar to what \citet{Meena2025} found (a more in-depth review on this topic is given by  \citealt{Diego2024b}; we specifically use Eq. 17 therein). On the contrary, if we fix $\beta=2.5$, we found that the mean value of the surface mass density of microlenses is $\Sigma_\ast=104\pm21$~M$_\odot\mathrm{pc}^{-2}$, twice as large as derived from intracluster light measurements~\citep{Montes2022}. Note that these measurements are IMF-dependent, and a Salpeter IMF would yield approximately twice the mass of a Chabrier IMF for the same light output. We expect that regions near a cluster member would show an enhanced density.

\begin{figure*} 
  \includegraphics[width=\linewidth]{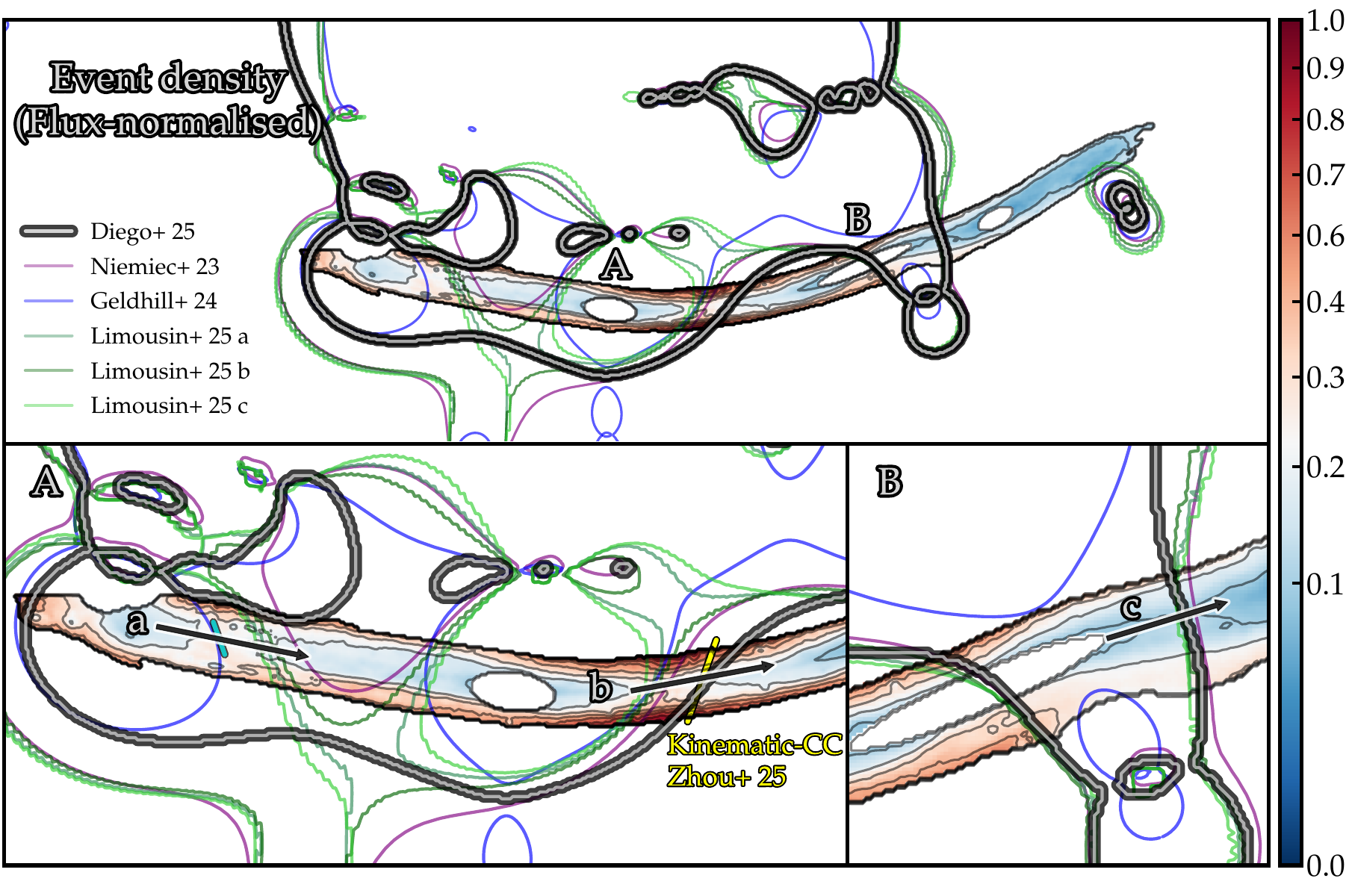} 
\caption{Event density distribution across the Dragon arc, normalised by the arc flux and rescaled to unity. The density is shown within a mask enclosing the arc to avoid artificially enhanced values from low background fluxes. Maxima trace the location of the CCs, and are compared with several lens models of Abell~370: the fiducial \citet{Diego2025} model (thick grey), and \citet{Limousin2025}, \citet{Gledhill2024}, and \citet{Niemiec2023} (green, blue, and purple, respectively), and a kinematic-CC model (yellow) \citep{Zhou2025}. The second row shows zoom-ins of regions A and B. Regions $a$, $b$, and $c$ mark directions used to analyse the event density along the principal stretching axes, with the cyan segment indicating a potential CC crossing. Maxima near the edges of the mask arise from low flux values in the normalisation and are therefore biased; edge regions are excluded.}
         \label{fig:normalised_event_density}
\end{figure*}

\begin{figure*} 
  \includegraphics[width=\linewidth]{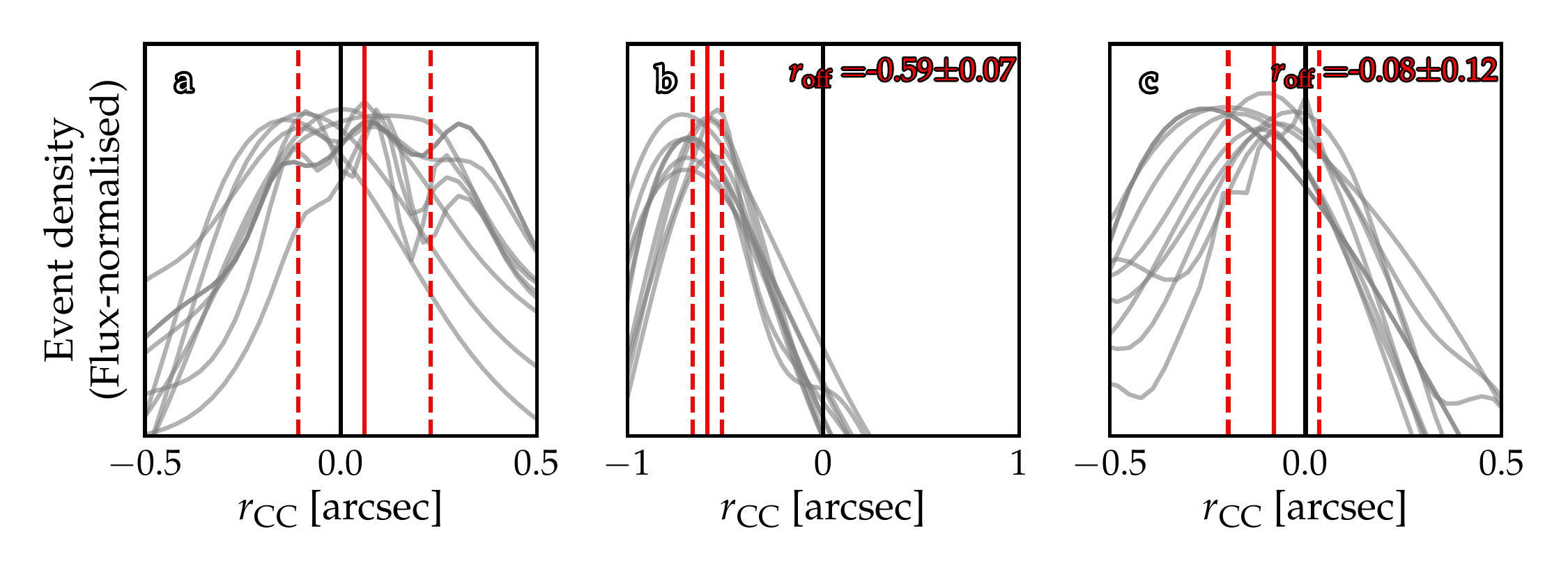} 
\caption{Flux-normalised event density along the trajectories $a$, $b$, and $c$ shown in Fig.~\ref{fig:normalised_event_density}. For each trajectory, we consider ten different paths with the same direction along the stretching of the arc, obtained by increasing or decreasing the $y$-pixel indices by 2 from curve to curve. The resulting profiles are shown as light-grey curves. The solid red line marks the mean position of the maxima of the curves, while the dashed lines indicate the corresponding $1\sigma$ dispersion. The solid black line denotes the position of the CC for the \citet{Diego2025} model, with the exception of trajectory $a$ (left panel), where the model does not predict a CC crossing the arc. In the latter case, we compare the results with a hypothetical CC crossing the arc at the position indicated by the cyan segment in Fig.~\ref{fig:normalised_event_density}. The trajectories are taken from east to west (left to right).}

         \label{fig:CC_trace}
\end{figure*}

\subsection{Improvement of the lens models}

The event locations can be used to trace regions of high magnification along the arc (i.e., the CC), assuming $\beta>2$. Earlier results suggested $\beta\approx2.5$ for this arc \citep{Diego2024b,Broadhurst2025}, while we find $\beta=2.18^{+0.20}_{-0.30}$ (consistent within $1\,\sigma$). Hence, regardless of the assumed DM model, the maximum number of events is expected near the CC for luminosity functions with $\beta>2$, even at high microlens densities where the ``more is less'' effect kicks in. This leads to two nearby maxima on opposite sides of the CC, enclosing a local minimum at its position.

After accounting for the arc flux, the event distribution can therefore be used to identify the expected location of the CC. To do this, we use the event density map normalised by the arc flux. For $\beta>2$, the bulk of events cluster both in regions of high magnification and in regions hosting a large number of potentially microlensed sources. Dividing by the arc flux corrects for the latter effect, as brighter regions contain larger stellar populations, yielding a map that depends primarily on magnification. This traces the CCs and any substructures that locally enhance the magnification, such as millilenses. While this normalisation by the continuum surface brightness provides a reasonable approximation, a normalisation based on star formation tracers, such as $H_\alpha$, would likely yield improved results in future analyses.

We construct this map by stacking all epochs for each filter, thereby removing transient events and obtaining a clean image of the arc in each band. We then compute the pixel-by-pixel mean across the stacked images and divide the event density map by this value. The resulting weighted event density map accounts for variations in surface brightness, correcting for regions with higher stellar densities and therefore higher event probabilities. This approach is lens model independent and provides an unbiased tracer of the microlensing probability. While this probability correlates with magnification, it is slightly biased towards the negative parity side of the CC, where more events are expected~\citep{Palencia2024}. As a result, the true CC should be slightly offset from the maxima in the weighted event density map. The magnitude of this offset is expected to scale with the microlens surface mass density, $\Sigma_*$ \citep{Diego2024b}.

In Fig.~\ref{fig:normalised_event_density} we show this weighted event density map, normalising the event values to a range 0--1, as a blue to red colour map, where the redder colours indicate the areas in which a higher number of events is expected as taken from the data, thus the maxima locate the positions of the CCs. Maxima near the edges of the mask arise from low flux values in the normalisation, which artificially boosts the signal when dividing by the flux. As a result, these regions are biased and are excluded from the analysis. The lower panels in Fig.~\ref{fig:normalised_event_density} show two zoom-in views of locations A and B in the top panel. Focusing on region B, we expect two crossings of the CC from the \citet{Diego2025} model. We can see that two maxima appear almost exactly at the position of the CC predicted by the model, shown as a thick grey line. Other models such as those from \citet{Limousin2025}, \citet{Gledhill2024}, and \citet{Niemiec2023} are respectively displayed in green, blue, and purple, where most models seem to agree. Focusing on region A, we can see two maxima. The maximum at the western position (right) shows some dispersion, but again most models trace well the position of the CC. Finally, on the eastern maximum, we see a broad band rather than a narrow peak. This could be either from a crossing of the CC, as predicted by most models, or from the proximity of a CC that does not cross the arc completely, as suggested by \citet{Diego2025}. The model by \citet{Gledhill2024} seems to be the closest one to the maximum on the eastern ``crossing,'' but this model, and all the others except the one from \citet{Diego2025}, predict a second crossing in the central part of region A, where no maximum is located. 

To better quantify this, we have taken trajectories along the direction of stretch of the arc, at the possible crossings $a$, $b$, and $c$ in the bottom panels of Fig.~\ref{fig:normalised_event_density}. We test against the model of \citet{Diego2025}. Fig.~\ref{fig:CC_trace} shows in grey the trajectories along the arc direction, separated by 2 pixels in the ordinate axis, for 10 paths, and their distance to the CC predicted by the model. In the case of the trajectories in $a$, we use a hypothetical CC indicated by a cyan segment in Fig.~\ref{fig:normalised_event_density}. The thick red vertical lines show the mean distance of the maxima from the model guess, and the dashed lines indicate the location at one standard deviation from the mean. For the hypothetical CC, the mean offset from the CC is well within the 1 $\sigma$ contour. Case $b$ shows the largest offset, with a mean offset of $0.59''$, corresponding to a 7.9$\sigma$ tension with the data. This tension appears only for the fiducial lens model~\citep{Diego2025}, while other models predict a CC much closer to the maximum at $b$ but also predict an additional CC crossing between $a$ and $b$, which is not supported by the event tracer method. Case $b$ is also offset from kinematic-based predictions for the CC crossing~\citep{Zhou2025} in Fig.~\ref{fig:normalised_event_density}, shown as a yellow line, which instead agrees better with our fiducial model. Finally, the paths along the $c$ trajectory show a mean offset of $0.08''$, or 0.7$\sigma$ from the model prediction. This work shows that a large quantity of CCEs can be used to trace the position of the CC crossing giant arcs, provided that $\beta>2$. Offsets between the event-traced CC and lens model predictions can be used to refine the models, or may instead reflect additional effects such as unresolved substructure or variations in the underlying stellar population (e.g. regions of enhanced star formation).

\section{Conclusion} \label{sec:conclusion}
In this work we have carried out a systematic search for CCEs in the giant arc ``the Dragon'' behind Abell~370, using multi-epoch JWST/NIRCam imaging in four filters (F150W, F182M, F200W, and F210M). By constructing empirical PSFs for each epoch, drizzling all data to a common astrometric frame, and adopting a residual-analysis technique in which each epoch is compared to the median of the remaining ones, we have identified transient, point-like flux variations along the arc. The significance of these events is calibrated empirically using the Dragon's head, an extra image of the background source with similar surface brightness to the arc, providing a robust characterisation of the intrinsic fluctuation field. Combining detections from {\tt DAOFIND} and {\tt SExtractor}, and accounting for repeated detections across filters and epochs, we obtain a catalogue of 104 unique CCEs in the Dragon arc, which currently represents the largest sample of highly magnified star candidates  identified in a single galaxy.

The size of this sample allows us to begin exploring the properties of the underlying stellar population at $z \approx 0.725$. A simple colour–magnitude analysis of events detected at high significance in at least two filters shows a roughly equal number of blue and red stars, with a small fraction of yellow candidates. This suggests that both BSGs and RSGs  contribute to the observed CCEs, in agreement with expectations for a relatively evolved, still star-forming system. 

We also investigate the implications of the spatial distribution of events for DM models. The parity-dependent (for our fiducial model) event statistics along the arc show similar behaviour at low to intermediate macromagnification values, but reveal a deficit of events near the CC on the positive-parity side, together with a persistent population and a mild excess at intermediate macromagnification on the negative-parity side. This pattern is qualitatively consistent with expectations in FDM scenarios, where surface-density fluctuations can enhance the probability of CCEs on the negative-parity side of the CC \citep{Broadhurst2025}. This picture is also consistent with a standard cold DM scenario, where even with only stellar microlenses more events are predicted on the negative-parity side of the CC \citep{Venumadhav2017,Palencia2024}. At the same time, modest shifts in the CC location in the macroscopic lens model could reproduce similar trends. At present, the data therefore provide an interesting hint, but not a decisive test, and improved modelling together with larger samples of arcs will be required to robustly separate DM-driven effects from residual lens-model systematics. When considering model-independent CC estimates based on symmetries, as \citet{Broadhurst2025} did, the asymmetry is still present and appears larger than expected from microlensing alone. To confirm this possible indication of an FDM scenario, a more detailed analysis must be performed using the newly identified events.

Finally, we demonstrate that the flux-normalised event density can be used as an empirical tracer of the CC. By constructing an event-density map weighted by the local surface brightness of the arc, we recover well-defined ridges that closely follow the expected position of the CC. In region~B, the maxima are in excellent agreement with the CC predicted by the free-form model of \citet{Diego2025}, as well as with several independent parametric models. In region~A, our event-based tracer confirms one of the predicted crossings, although the model maximum shows a larger offset from the data compared to other crossings. The data show a hint of a second crossing in the eastern part of the arc (region A-$a$), where a maximum is observed, but whether this comes from a CC crossing or a nearby CC boosting the local macromagnification remains uncertain. A more quantitative comparison along trajectories aligned with the arc shows that the model of \citet{Diego2025} is fully consistent with the data in all of the three tested regions; however, in the central part (A-$b$) the event distribution favours a CC location closer to that predicted by alternative models and offsets from our fiducial and the kinematic driven constraint. This highlights the potential of large CCE samples as an additional, independent set of observables to refine lens models locally along bright arcs.

Altogether, the Dragon arc provides a first example of how multi-epoch JWST imaging can deliver sizeable samples of highly magnified stars in a single system, enabling joint studies of stellar populations, DM physics, and the fine structure of cluster CCs. Extending this analysis to other strongly lensed galaxies, and combining it with higher-cadence monitoring and improved lens modelling, will turn microlensed stars from rare curiosities into a routine probe of both baryons and DM in and around massive galaxy clusters.

\begin{acknowledgements}

This research is based on observations made with the NASA/ESA James Webb Space Telescope, obtained from the Space Telescope Science Institute (STScI), which is operated by the Association of Universities for Research in Astronomy, Inc., under NASA contract NAS 5–26555. These observations are associated with programs GTO-1208, GO-2883, GO-3538, and GO-5324. The data are available at MAST: \url{https://mast.stsci.edu/portal/Mashup/Clients/Mast/Portal.html}

J.M.P. acknowledges financial support from the Formaci\'on de Personal Investigador (FPI) programme, ref. PRE2020-096261, associated with the Spanish Agencia Estatal de Investigaci\'on project MDM-2017-0765-20-2, partial financial support from the Complementary Plan in Astrophysics and High-Energy Physics (CA25944), project C17.I02.P02.S01.S03 CSIC, supported by the Next Generation EU funds, RRF and PRTR mechanisms, and the Government of the Autonomous Community of Cantabria, and acknowledges funding from the project UC-LIME (PID2022-140670NA-I00), financed by MCIN/AEI/ 10.13039/501100011033/FEDER, UE.
J.M.P. and J.M.D. acknowledge the support of projects PID2022-138896NB-C51 (MCIU/AEI/MINECO/FEDER, UE) Ministerio de Ciencia, Investigaci\'on y Universidades and SA101P24. 
A.V.F. acknowledges the support of many individual donors.
A.Z. acknowledges support by the Israel Science Foundation Grant No. 864/23.
D.E. acknowledges support from projects PID2023-150178NB-I00 and PID2023-149578NB-I00,  financed by MCIN/AEI/10.13039/501100011033, and by FEDER, EU.
C.N.A.W. acknowledges funding from JWST/NIRCam contract to the University of Arizona NAS5-02105
S.K.L. acknowledge support from the Research Grants Council (RGC) of Hong Kong through the General Research Fund (GRF) 17312122.
P.M. and T.J.B. are supported by the Spanish Grant PID2023-149016NB-I00 (MINECO/AEI/FEDER,UE). PM also acknowledges financial support from fellowship PIF22/177 (UPV/EHU).
E.Z. acknowledges project grant 2022-03804 from the Swedish Research Council (Vetenskapsr\aa{}det).

We made use of the following software: Python, NumPy, SciPy, Matplotlib, Astropy, SExtractor, Photutils, DAOFIND.
\end{acknowledgements}
 
\bibliography{bibtex}{}
\bibliographystyle{aasjournal}

\appendix

\onecolumn
\section{Catalogue of Events}\label{app:catalogue}

\begin{longtable}{ccccccccc}
\caption{Reported transient events in the Dragon arc. Events detected at approximately the same sky position in consecutive epochs are treated as a single event and are therefore assigned the same identification number. For each event, the table reports the J2000 sky coordinates (RA and DEC), the detection significance in each filter at the epoch of detection, the method used for the identification, and the date of the corresponding observational epoch.
\label{tab:catalogue}}\\
\hline\hline
ID & RA (deg) & Dec. (deg) & $\sigma_{\mathrm{F150W}}$ & $\sigma_{\mathrm{F182M}}$ & $\sigma_{\mathrm{F200W}}$ & $\sigma_{\mathrm{F210M}}$ & Method$^{b}$ & Date
\\
\hline
\endfirsthead
\caption{Continued on next page.}\\
\hline\hline
ID & RA (deg) & DEC (deg) & $\sigma_{\mathrm{F150W}}$ & $\sigma_{\mathrm{F182M}}$ & $\sigma_{\mathrm{F200W}}$ & $\sigma_{\mathrm{F210M}}$ & Method$^{b}$ & Date \\
\hline
\endhead
\hline
\endfoot
\hline
$1^a$ & 39.9722445 & $-$1.5846997 & 3.72 & - & 8.18 & - & {	\tt DAOFIND} & 2024-09-09 \\
$1^a$ & 39.9722445 & $-$1.5847080 & 3.47 & - & 6.24 & - & {	\tt DAOFIND} & 2024-12-17 \\
$2^a$ & 39.9690183 & $-$1.5846663 & - & - & - & 6.00 & {	\tt DAOFIND} & 2024-07-25 \\
$3^a$ & 39.9690183 & $-$1.5846663 & 4.09 & - & 6.68 & - & {	\tt DAOFIND} & 2024-12-17 \\
$4^a$ & 39.9720194 & $-$1.5847830 & - & - & - & 14.75 & {	\tt DAOFIND} & 2024-08-02 \\
$4^a$ & 39.9720194 & $-$1.5847830 & - & - & - & 12.08 & {	\tt DAOFIND} & 2024-07-25 \\
$4^a$ & 39.9720194 & $-$1.5847830 & - & 16.58 & - & - & {	\tt DAOFIND} & 2024-07-31 \\
$4^a$ & 39.9720445 & $-$1.5847830 & 1.93 & - & 7.78 & - & {	\tt DAOFIND} & 2024-09-09 \\
$4^a$ & 39.9720361 & $-$1.5847830 & 5.64 & - & 5.21 & - & {	\tt SExtractor} & 2024-12-17 \\
$5^a$ & 39.9719528 & $-$1.5848413 & 6.48 & - & 7.96 & - & {	\tt DAOFIND} & 2024-12-17 \\
$6^{\,\,\,}$ & 39.9727781 & $-$1.5847330 & 2.72 & - & 8.16 & - & {	\tt DAOFIND} & 2024-09-09 \\
$6^{\,\,\,}$ & 39.9727864 & $-$1.5847330 & 5.81 & - & 8.50 & - & {	\tt DAOFIND} & 2024-12-17 \\
$7^{\,\,\,}$ & 39.9693518 & $-$1.5847580 & 5.60 & - & 3.49 & - & {	\tt DAOFIND} & 2022-12-29 \\
$8^{\,\,\,}$ & 39.9707356 & $-$1.5850413 & 5.70 & - & 6.90 & - & {	\tt DAOFIND} & 2022-12-29 \\
$9^{\,\,\,}$ & 39.9707189 & $-$1.5850330 & - & 5.53 & - & - & {	\tt SExtractor} & 2024-07-31 \\
$10^a$ & 39.9689933 & $-$1.5846496 & 6.17 & - & 7.81 & - & {	\tt DAOFIND} & 2022-12-29 \\
$11^{\,\,\,}$ & 39.9719944 & $-$1.5848080 & 6.30 & - & 12.83 & - & {	\tt DAOFIND} & 2022-12-29 \\
$11^{\,\,\,}$ & 39.9720111 & $-$1.5848163 & - & 3.71 & - & 5.21 & {	\tt SExtractor} & 2023-12-20 \\
$11^{\,\,\,}$ & 39.9720111 & $-$1.5848163 & - & 2.05 & - & 5.09 & {	\tt SExtractor} & 2023-12-19 \\
$12^a$ & 39.9727614 & $-$1.5847997 & 6.41 & - & 6.11 & - & {	\tt DAOFIND} & 2022-12-29 \\
$13^{\,\,\,}$ & 39.9694601 & $-$1.5847663 & 7.03 & - & 9.57 & - & {	\tt DAOFIND} & 2022-12-29 \\
$14^a$ & 39.9694101 & $-$1.5847830 & 7.07 & - & 2.83 & - & {	\tt DAOFIND} & 2022-12-29 \\
$15^a$ & 39.9727614 & $-$1.5847580 & 7.97 & - & 18.05 & - & {	\tt DAOFIND} & 2022-12-29 \\
$16^{\,\,\,}$ & 39.9727614 & $-$1.5847580 & - & - & - & 12.22 & {	\tt DAOFIND} & 2024-08-02 \\
$16^{\,\,\,}$ & 39.9727614 & $-$1.5847580 & - & 13.99 & - & - & {	\tt DAOFIND} & 2024-07-31 \\
$17^a$ & 39.9703771 & $-$1.5850830 & 9.31 & - & 7.78 & - & {	\tt DAOFIND} & 2022-12-29 \\
$17^a$ & 39.9703938 & $-$1.5850913 & - & 4.72 & - & 8.65 & {	\tt DAOFIND} & 2023-12-22 \\
$17^a$ & 39.9703938 & $-$1.5850913 & - & 7.55 & - & 6.73 & {	\tt DAOFIND} & 2023-12-20 \\
$17^a$ & 39.9703938 & $-$1.5850830 & - & 7.65 & - & 8.13 & {	\tt DAOFIND} & 2023-12-19 \\
$18^a$ & 39.9703938 & $-$1.5850913 & 12.60 & - & 11.99 & - & {	\tt DAOFIND} & 2024-12-17 \\
$19^{\,\,\,}$ & 39.9706189 & $-$1.5851497 & 17.04 & - & 21.57 & - & {	\tt DAOFIND} & 2022-12-29 \\
$19^{\,\,\,}$ & 39.9706189 & $-$1.5851413 & - & 9.32 & - & 8.63 & {	\tt DAOFIND} & 2023-12-22 \\
$19^{\,\,\,}$ & 39.9706272 & $-$1.5851497 & - & 7.42 & - & 11.14 & {	\tt DAOFIND} & 2023-12-20 \\
$19^{\,\,\,}$ & 39.9706189 & $-$1.5851497 & - & 9.46 & - & 10.92 & {	\tt DAOFIND} & 2023-12-19 \\
$20^{\,\,\,}$ & 39.9695102 & $-$1.5848163 & 5.15 & - & 1.08 & - & {	\tt DAOFIND} & 2024-12-17 \\
$21^{\,\,\,}$ & 39.9694601 & $-$1.5848163 & - & 5.36 & - & 2.85 & {	\tt DAOFIND} & 2023-12-19 \\
$22^{\,\,\,}$ & 39.9700354 & $-$1.5849496 & 2.07 & - & 7.05 & - & {	\tt DAOFIND} & 2022-12-29 \\
$22^{\,\,\,}$ & 39.9700354 & $-$1.5849580 & - & 6.52 & - & 6.11 & {	\tt DAOFIND} & 2023-12-20 \\
$22^{\,\,\,}$ & 39.9700520 & $-$1.5849746 & - & 6.72 & - & 9.78 & {	\tt SExtractor} & 2023-12-19 \\
$23^{\,\,\,}$ & 39.9707940 & $-$1.5850830 & 4.45 & - & 5.93 & - & {	\tt DAOFIND} & 2022-12-29 \\
$23^{\,\,\,}$ & 39.9708106 & $-$1.5850913 & - & 8.07 & - & 1.08 & {	\tt DAOFIND} & 2023-12-19 \\
$24^{\,\,\,}$ & 39.9703271 & $-$1.5849663 & - & 9.87 & - & 4.87 & {	\tt DAOFIND} & 2023-12-22 \\
$24^{\,\,\,}$ & 39.9703271 & $-$1.5849747 & - & 8.28 & - & 3.70 & {	\tt DAOFIND} & 2023-12-20 \\
$24^{\,\,\,}$ & 39.9703271 & $-$1.5849663 & - & 11.10 & - & 6.26 & {	\tt DAOFIND} & 2023-12-19 \\
$25^{\,\,\,}$ & 39.9711608 & $-$1.5850913 & - & 6.36 & - & 1.78 & {	\tt SExtractor} & 2023-12-19 \\
$26^{\,\,\,}$ & 39.9712275 & $-$1.5850330 & - & 5.20 & - & 1.60 & {	\tt SExtractor} & 2023-12-19 \\
$27^a$ & 39.9714692 & $-$1.5849913 & - & 5.01 & - & 3.09 & {	\tt SExtractor} & 2023-12-19 \\
$28^{\,\,\,}$ & 39.9704188 & $-$1.5851080 & 3.88 & - & 6.79 & - & {	\tt DAOFIND} & 2024-09-09 \\
$29^{\,\,\,}$ & 39.9703271 & $-$1.5849830 & - & 11.10 & - & 6.26 & {	\tt DAOFIND} & 2023-12-19 \\
$30^{\,\,\,}$ & 39.9696435 & $-$1.5847746 & - & 2.20 & - & 6.50 & {	\tt DAOFIND} & 2023-12-20 \\
$31^{\,\,\,}$ & 39.9695435 & $-$1.5848580 & - & 5.38 & - & 1.61 & {	\tt DAOFIND} & 2023-12-20 \\
$32^a$ & 39.9718527 & $-$1.5850163 & 3.90 & - & 7.62 & - & {	\tt DAOFIND} & 2022-12-29 \\
$32^a$ & 39.9718527 & $-$1.5850163 & - & 4.06 & - & 7.28 & {	\tt DAOFIND} & 2023-12-20 \\
$32^a$ & 39.9718444 & $-$1.5850163 & - & 3.80 & - & 7.87 & {	\tt DAOFIND} & 2023-12-19 \\
$33^{\,\,\,}$ & 39.9694351 & $-$1.5847663 & - & 5.62 & - & - & {	\tt DAOFIND} & 2024-07-31 \\
$34^{\,\,\,}$ & 39.9716110 & $-$1.5849497 & - & 5.70 & - & - & {	\tt DAOFIND} & 2024-07-31 \\
$35^a$ & 39.9713775 & $-$1.5849997 & - & 5.86 & - & - & {	\tt DAOFIND} & 2024-07-31 \\
$36^{\,\,\,}$ & 39.9704188 & $-$1.5849997 & - & 5.87 & - & - & {	\tt DAOFIND} & 2024-07-31 \\
$37^a$ & 39.9725447 & $-$1.5847664 & - & 6.02 & - & - & {	\tt DAOFIND} & 2024-07-31 \\
$38^{\,\,\,}$ & 39.9714359 & $-$1.5848330 & - & - & - & 6.83 & {	\tt DAOFIND} & 2024-08-02 \\
$38^{\,\,\,}$ & 39.9714276 & $-$1.5848330 & - & 6.28 & - & - & {	\tt DAOFIND} & 2024-07-31 \\
$39^{\,\,\,}$ & 39.9714276 & $-$1.5849913 & - & 6.58 & - & - & {	\tt DAOFIND} & 2024-07-31 \\
$40^{\,\,\,}$ & 39.9720611 & $-$1.5848163 & - & - & - & 5.93 & {	\tt DAOFIND} & 2024-07-25 \\
$40^{\,\,\,}$ & 39.9720528 & $-$1.5848163 & - & 7.64 & - & - & {	\tt DAOFIND} & 2024-07-31 \\
$41^{\,\,\,}$ & 39.9701187 & $-$1.5849580 & - & - & - & 6.00 & {	\tt DAOFIND} & 2024-07-25 \\
$41^{\,\,\,}$ & 39.9701187 & $-$1.5849496 & - & 7.80 & - & - & {	\tt DAOFIND} & 2024-07-31 \\
$42^a$ & 39.9691017 & $-$1.5847913 & - & 7.96 & - & - & {	\tt DAOFIND} & 2024-07-31 \\
$43^{\,\,\,}$ & 39.9721612 & $-$1.5847914 & - & - & - & 7.18 & {	\tt DAOFIND} & 2024-08-02 \\
$43^{\,\,\,}$ & 39.9721695 & $-$1.5847830 & - & - & - & 5.97 & {	\tt DAOFIND} & 2024-07-25 \\
$43^{\,\,\,}$ & 39.9721695 & $-$1.5847830 & - & 8.43 & - & - & {	\tt DAOFIND} & 2024-07-31 \\
$44^{\,\,\,}$ & 39.9694851 & $-$1.5848080 & - & - & - & 7.10 & {	\tt DAOFIND} & 2024-07-25 \\
$44^{\,\,\,}$ & 39.9694935 & $-$1.5847996 & - & 8.92 & - & - & {	\tt DAOFIND} & 2024-07-31 \\
$45^{\,\,\,}$ & 39.9723529 & $-$1.5847747 & - & 9.27 & - & - & {	\tt DAOFIND} & 2024-07-31 \\
$46^{\,\,\,}$ & 39.9704438 & $-$1.5850997 & - & - & - & 5.80 & {	\tt DAOFIND} & 2024-07-25 \\
$46^{\,\,\,}$ & 39.9704522 & $-$1.5850913 & - & 9.48 & - & - & {	\tt DAOFIND} & 2024-07-31 \\
$47^{\,\,\,}$ & 39.9708440 & $-$1.5850497 & - & 9.75 & - & - & {	\tt DAOFIND} & 2024-07-31 \\
$48^a$ & 39.9706439 & $-$1.5851413 & - & 5.21 & - & - & {	\tt SExtractor} & 2024-07-31 \\
$49^a$ & 39.9706272 & $-$1.5851497 & - & 7.42 & - & 11.14 & {	\tt DAOFIND} & 2023-12-20 \\
$50^{\,\,\,}$ & 39.9703771 & $-$1.5851080 & - & - & - & 10.15 & {	\tt SExtractor} & 2024-08-02 \\
$50^{\,\,\,}$ & 39.9703688 & $-$1.5851080 & - & 6.29 & - & - & {	\tt SExtractor} & 2024-07-31 \\
$51^{\,\,\,}$ & 39.9715609 & $-$1.5850080 & - & 5.69 & - & - & {	\tt SExtractor} & 2024-07-31 \\
$52^{\,\,\,}$ & 39.9703021 & $-$1.5849580 & - & 6.27 & - & - & {	\tt SExtractor} & 2024-07-31 \\
$53^{\,\,\,}$ & 39.9703438 & $-$1.5849497 & - & - & - & 6.36 & {	\tt SExtractor} & 2024-08-02 \\
$53^{\,\,\,}$ & 39.9703521 & $-$1.5849497 & - & 7.83 & - & - & {	\tt SExtractor} & 2024-07-31 \\
$54^a$ & 39.9688599 & $-$1.5845496 & - & 5.08 & - & - & {	\tt SExtractor} & 2024-07-31 \\
$55^a$ & 39.9700687 & $-$1.5849163 & 3.54 & - & 7.58 & - & {	\tt DAOFIND} & 2022-12-29 \\
$56^a$ & 39.9700687 & $-$1.5849163 & - & - & - & 9.45 & {	\tt DAOFIND} & 2024-08-02 \\
$57^{\,\,\,}$ & 39.9700687 & $-$1.5849163 & - & - & - & 11.35 & {	\tt DAOFIND} & 2024-07-25 \\
$58^a$ & 39.9698853 & $-$1.5849830 & - & 6.01 & - & 2.15 & {	\tt DAOFIND} & 2023-12-20 \\
$59^a$ & 39.9700520 & $-$1.5849746 & - & 6.72 & - & 9.78 & {	\tt SExtractor} & 2023-12-19 \\
$60^{\,\,\,}$ & 39.9694685 & $-$1.5847330 & 4.49 & - & 5.21 & - & {	\tt SExtractor} & 2024-12-17 \\
$61^{\,\,\,}$ & 39.9718694 & $-$1.5847330 & 3.01 & - & 5.71 & - & {	\tt DAOFIND} & 2022-12-29 \\
$62^{\,\,\,}$ & 39.9695018 & $-$1.5847746 & 2.79 & - & 6.17 & - & {	\tt DAOFIND} & 2022-12-29 \\
$63^{\,\,\,}$ & 39.9696852 & $-$1.5847996 & 1.84 & - & 6.18 & - & {	\tt DAOFIND} & 2022-12-29 \\
$64^{\,\,\,}$ & 39.9692684 & $-$1.5847580 & 3.22 & - & 6.24 & - & {	\tt DAOFIND} & 2022-12-29 \\
$65^{\,\,\,}$ & 39.9712525 & $-$1.5850580 & 2.30 & - & 6.41 & - & {	\tt DAOFIND} & 2022-12-29 \\
$66^{\,\,\,}$ & 39.9717193 & $-$1.5848497 & 3.92 & - & 6.48 & - & {	\tt DAOFIND} & 2022-12-29 \\
$67^a$ & 39.9720361 & $-$1.5848330 & 3.43 & - & 6.80 & - & {	\tt DAOFIND} & 2022-12-29 \\
$68^a$ & 39.9704438 & $-$1.5850580 & 2.66 & - & 7.02 & - & {	\tt DAOFIND} & 2022-12-29 \\
$69^{\,\,\,}$ & 39.9704438 & $-$1.5850497 & - & - & - & 6.36 & {	\tt DAOFIND} & 2024-08-02 \\
$70^{\,\,\,}$ & 39.9720945 & $-$1.5848247 & 3.45 & - & 7.61 & - & {	\tt DAOFIND} & 2022-12-29 \\
$71^{\,\,\,}$ & 39.9717610 & $-$1.5848413 & 3.36 & - & 7.76 & - & {	\tt DAOFIND} & 2022-12-29 \\
$72^{\,\,\,}$ & 39.9704105 & $-$1.5849497 & 5.68 & - & 7.87 & - & {	\tt DAOFIND} & 2022-12-29 \\
$73^a$ & 39.9704438 & $-$1.5850330 & 3.33 & - & 7.95 & - & {	\tt DAOFIND} & 2022-12-29 \\
$74^{\,\,\,}$ & 39.9704438 & $-$1.5850497 & - & - & - & 6.36 & {	\tt DAOFIND} & 2024-08-02 \\
$75^{\,\,\,}$ & 39.9722029 & $-$1.5847497 & 2.81 & - & 5.39 & - & {	\tt SExtractor} & 2022-12-29 \\
$76^a$ & 39.9704438 & $-$1.5850997 & - & - & - & 5.80 & {	\tt DAOFIND} & 2024-07-25 \\
$77^{\,\,\,}$ & 39.9704188 & $-$1.5851080 & 3.88 & - & 6.79 & - & {	\tt DAOFIND} & 2024-09-09 \\
$78^{\,\,\,}$ & 39.9711775 & $-$1.5850413 & 1.66 & - & 6.00 & - & {	\tt DAOFIND} & 2024-12-17 \\
$79^{\,\,\,}$ & 39.9707523 & $-$1.5850580 & 3.32 & - & 6.41 & - & {	\tt DAOFIND} & 2024-12-17 \\
$80^{\,\,\,}$ & 39.9708273 & $-$1.5850663 & 2.36 & - & 5.89 & - & {	\tt DAOFIND} & 2024-09-09 \\
$81^{\,\,\,}$ & 39.9705355 & $-$1.5851080 & 2.04 & - & 5.90 & - & {	\tt DAOFIND} & 2024-09-09 \\
$82^a$ & 39.9718027 & $-$1.5849497 & 2.64 & - & 6.15 & - & {	\tt DAOFIND} & 2024-09-09 \\
$83^a$ & 39.9720111 & $-$1.5848580 & - & - & - & 6.80 & {	\tt DAOFIND} & 2024-07-25 \\
$84^a$ & 39.9703772 & $-$1.5849997 & - & 2.25 & - & 5.03 & {	\tt SExtractor} & 2023-12-19 \\
$85^{\,\,\,}$ & 39.9720194 & $-$1.5847830 & - & - & - & 14.75 & {	\tt DAOFIND} & 2024-08-02 \\
$85^{\,\,\,}$ & 39.9720194 & $-$1.5847830 & - & - & - & 12.08 & {	\tt DAOFIND} & 2024-07-25 \\
$85^{\,\,\,}$ & 39.9720194 & $-$1.5847830 & - & 16.58 & - & - & {	\tt DAOFIND} & 2024-07-31 \\
$86^{\,\,\,}$ & 39.9710357 & $-$1.5849997 & - & 2.45 & - & 5.94 & {	\tt DAOFIND} & 2023-12-22 \\
$87^{\,\,\,}$ & 39.9702438 & $-$1.5849913 & - & 2.36 & - & 6.23 & {	\tt DAOFIND} & 2023-12-22 \\
$88^{\,\,\,}$ & 39.9702271 & $-$1.5850080 & - & 2.12 & - & 5.73 & {	\tt DAOFIND} & 2023-12-20 \\
$89^{\,\,\,}$ & 39.9699687 & $-$1.5849663 & - & 3.30 & - & 5.81 & {	\tt DAOFIND} & 2023-12-20 \\
$90^{\,\,\,}$ & 39.9705022 & $-$1.5851080 & - & 2.70 & - & 5.90 & {	\tt DAOFIND} & 2023-12-20 \\
$91^{\,\,\,}$ & 39.9712025 & $-$1.5850080 & - & 2.22 & - & 6.48 & {	\tt DAOFIND} & 2023-12-20 \\
$92^a$ & 39.9703855 & $-$1.5850830 & 9.31 & - & 7.78 & - & {	\tt SExtractor} & 2022-12-29 \\
$92^a$ & 39.9703938 & $-$1.5850913 & - & 4.72 & - & 8.65 & {	\tt DAOFIND} & 2023-12-22 \\
$92^a$ & 39.9703938 & $-$1.5850913 & - & 7.55 & - & 6.73 & {	\tt DAOFIND} & 2023-12-20 \\
$92^a$ & 39.9703938 & $-$1.5850830 & - & 7.65 & - & 8.13 & {	\tt DAOFIND} & 2023-12-19 \\
$93^{\,\,\,}$ & 39.9703938 & $-$1.5850913 & 12.60 & - & 11.99 & - & {	\tt DAOFIND} & 2024-12-17 \\
$94^{\,\,\,}$ & 39.9685181 & $-$1.5844413 & - & 2.35 & - & 6.76 & {	\tt DAOFIND} & 2023-12-20 \\
$95^{\,\,\,}$ & 39.9690850 & $-$1.5846496 & - & 2.65 & - & 7.08 & {	\tt DAOFIND} & 2023-12-20 \\
$96^{\,\,\,}$ & 39.9705355 & $-$1.5850580 & - & 2.39 & - & 13.83 & {	\tt SExtractor} & 2023-12-20 \\
$97^{\,\,\,}$ & 39.9713108 & $-$1.5849913 & - & 4.98 & - & 5.00 & {	\tt SExtractor} & 2023-12-20 \\
$98^{\,\,\,}$ & 39.9695352 & $-$1.5848913 & - & - & - & 5.86 & {	\tt DAOFIND} & 2024-07-25 \\
$99^a$ & 39.9703938 & $-$1.5849747 & - & - & - & 6.68 & {	\tt DAOFIND} & 2024-07-25 \\
$100^a$ & 39.9714442 & $-$1.5849080 & - & - & - & 6.52 & {	\tt SExtractor} & 2024-07-25 \\
$101^{\,\,\,}$ & 39.9694101 & $-$1.5848163 & - & - & - & 5.82 & {	\tt DAOFIND} & 2024-08-02 \\
$102^{\,\,\,}$ & 39.9699853 & $-$1.5849913 & - & - & - & 5.99 & {	\tt DAOFIND} & 2024-08-02 \\
$103^{\,\,\,}$ & 39.9699770 & $-$1.5849330 & - & - & - & 6.12 & {	\tt DAOFIND} & 2024-08-02 \\
$104^{\,\,\,}$ & 39.9705105 & $-$1.5849913 & - & - & - & 5.22 & {	\tt SExtractor} & 2024-08-02 \\
\end{longtable}
{\bf Notes:}
$^{a}$Events at the positions of \citet{Fudamoto2025} events.
$^{b}${\tt SExtractor} significances are first estimated assuming independent pixels within the event mask and subsequently corrected for a systematic bias calibrated using events detected by both {\tt DAOFIND} and {\tt SExtractor}.

\end{document}